\documentclass[12pt,a4paper,twoside]{article}
\usepackage{graphics}
\usepackage{multirow}
\usepackage{fancyheadings}
\newlength{\figwidth}
\newlength{\figheight}
\newcommand{\bigint}[2]{\displaystyle \int\limits_{#1}^{#2}} 
\newcommand{\bigmod}[2]{\left.{#1}\strut\right|_{#2}}
\newcommand{\eqref}[1]{(\ref{#1})}
\begin{document}

\pagestyle{empty}

\begin{center}
  \large\sc 
  Institute of Experimental Physics \\
  Warsaw University
\end{center}

\vspace{1cm}
\begin{flushright}
\begin{tabular}{l}
 IFD - 02/99 \\
 hep-ph/9904334 \\
 DESY 99-074 \\[3mm]
 {\bf Euro. Phys. J. C} \\
 {\bf Vol. 11 (1999) 539}
\end{tabular}
\end{flushright}

\begin{center}

\vspace{1.5cm}
\Large
Aleksander Filip \.Zarnecki

\vspace{0.5cm}
\Large\bf
Global Analysis of $eeqq$ Contact Interactions 
and 
Future Prospects for High Energy Physics 

\vfill
\normalsize\rm
Warszawa, June 1999

\end{center}

\newpage

%----------------------------------------------------------------------

\vspace*{10cm}
\begin{abstract}
Data from HERA, LEP and the Tevatron, as well as from low energy  
experiments are used to constrain the scale of possible electron-quark 
contact interactions. Different models are considered, including the most 
general one, in which all new couplings can vary independently. Limits on 
couplings and mass scales are extracted and upper limits on 
possible effects to be observed in future HERA, LEP and Tevatron running
are estimated. 
Total hadronic cross-section at LEP and $e^{-}p$ scattering
cross-section at HERA are strongly constrained by existing data,
whereas large cross-section deviations are still possible for 
Drell-Yan lepton pair production at the Tevatron.
\end{abstract}

\newpage
% 
%---------------------------------------------------------------------------
%

\setcounter{page}{1}
\pagestyle{fancy}
\thispagestyle{plain}

\lhead[]{\fancyplain{}{\small A.F.\.Zarnecki}}
\rhead[\fancyplain{}{\small Global Analysis of $eeqq$ Contact %
 Interactions  and Future Prospects for High Energy Physics %
                     }]{\fancyplain{}{\sl\leftmark}}

%\lhead[\fancyplain{\sl\leftmark}{\sl\leftmark}]{}
%\rhead[]{\fancyplain{\sl\leftmark}{\sl\rightmark}}
% 
%---------------------------------------------------------------------------
%

\section{Introduction}
\label{sec-intro}

Search for "new physics" has always been one of the most exciting subjects
in the field of particle physics. The results presented in 1997 by 
the H1 \cite{h1_97} and ZEUS \cite{zeus97} experiments at HERA electrified the 
physics community. Both experiments reported  
an excess of events in positron-proton Neutral Current Deep 
Inelastic Scattering (NC DIS) at very high momentum transfer 
scales $Q^{2}$, as compared with the predictions of the Standard Model. 
Unfortunately, in spite of the significant increase in the integrated
data luminosity, these results have not been confirmed nor 
contradicted\cite{h1_98,zeus98}. The effect can be just due to a statistical
fluctuation, but can also be a first sign of some "new physics".

In 1998 HERA experiments started again\footnote{Previously HERA run in 
electron-proton mode in 1992-94.} to collect electron-proton data
aiming for integrated luminosity comparable with that of
the positron-proton data. The first results are expected soon.
The aim of the presented analysis is
to review experimental and theoretical constraints on possible signals
of "new physics" at HERA and extract limits on a new effects to be
seen in the  new HERA $e^-p$ data. 
Limits corresponding to other present and future high-energy
experiments are also considered. 
The contact interaction models, used as the general framework for this
analysis are described in  section \ref{sec-model}. 
In section \ref{sec-data} the relevant data 
from HERA, LEP, the Tevatron and other experiments are briefly 
described. Methods used to compare data with contact interaction 
model predictions are discussed in section \ref{sec-method}.
The results of analysis within different contact interaction 
models, including extracted limits on the mass scale of new interactions, 
are presented in section \ref{sec-results}. Predictions for 
the future discovery potential at HERA, as well as at LEP and the Tevatron 
are discussed in section~\ref{sec-predictions}.

The analysis presented here is based on the approach suggested in
\cite{ciglob}.  
Significant work has been done to improve the treatment of 
experimental data, including a proper interpretation of statistical 
and systematic errors as well as acceptance cuts and smearing. 

%---------------------------------------------------------------------------

\section{Contact Interactions}
\label{sec-model}

Four-fermion contact interactions are an effective theory, which 
allows us to describe, in the most general way, possible low energy 
effects  coming from "new physics" at much higher energy scales. This 
includes the possible existence of second generation heavy weak bosons, 
leptoquarks as well as electron and quark compositeness \cite{cidef,cihera}.
Contact interactions can be represented as additional terms in the 
Standard Model Lagrangian~\cite{cihera}:
\begin{eqnarray}
L_{CI} & = & \eta_{s}  (\bar{e}_{L} e_{R} )(\bar{q}_{L} q_{R}) +
\eta_{s}' (\bar{e}_{L} e_{R} )(\bar{q}_{R} q_{L}) + h.c. 
            \hspace{2.5cm} \rm scalar  \label{eq-lagr}\\[3mm]
       & + & \sum_{i,j=L,R} \eta_{ij} (\bar{e}_{i} \gamma^{\mu} e_{i} )
              (\bar{q}_{j} \gamma_{\mu} q_{j}) 
            \hspace{5.1cm} \rm vector \nonumber \\[2mm]
       & + & \eta_{T} (\bar{e}_{L} \sigma^{\mu \nu} e_{R} )
                      (\bar{q}_{L} \sigma_{\mu \nu} q_{R} ) + h.c.  
            \hspace{4.4cm} \rm tensor \nonumber
\end{eqnarray}
where subsequent lines describe the scalar, vector and tensor
contact interaction terms respectively. 
As very strong limits have been already placed on both scalar 
and tensor terms \cite{cihera} this paper considers vector terms only.

The influence of the vector contact interactions on the $ep$ NC DIS 
cross-section can be described as an additional term in the tree level 
$eq \rightarrow eq$ scattering amplitude~\cite{ciglob}:
\begin{eqnarray}
M^{e_i q_j \rightarrow e_i q_j}(t) & = & 
- \frac{4 \pi \alpha_{em} e_{q}}{t} \; + \;
  \frac{4 \pi \alpha_{em}}{sin^{2}\theta_{W} \cdot cos^{2}\theta_{W} } 
\cdot \frac{g^{e}_{i} g^{q}_{j}}{t - M_{Z}^{2}}
  \; + \; \eta^{eq}_{ij} \label{eq-mt}
\end{eqnarray}
where $t = -Q^{2}$ is the Mandelstam variable describing the
four-momentum transfer between the electron and the quark, $e_{q}$ is 
the electric charge of the quark in units of the elementary charge
and the subscripts $i$ and $j$ label the chiralities of the initial 
lepton and quark respectively: $i,j=L,R$.
$g^{e}_{i}$ and $g^{q}_{j}$ are electroweak 
couplings of the electron and the quark 
\begin{eqnarray}
g^{f}_{L} & = & I_{3f}     - e_{f} \sin^{2}\theta_{W} \label{eq-gdef} \\
g^{f}_{R} & = & \;\;\;\;\; - e_{f} \sin^{2}\theta_{W} \nonumber
\end{eqnarray}
where $I_{3f}$ is the third component of the SU(2) isospin for 
the fermion $f$: $f=e,q$.

For processes such as $e^{+}e^{-} \rightarrow hadrons$ or
$p \bar{p} \rightarrow l^{+} l^{-} X$, a corresponding formula 
can be written for the $e^{+}e^{-} \rightarrow q \bar{q}$ 
tree level amplitude:
\begin{eqnarray}
M^{e_i \bar{e}_j \rightarrow q_i \bar{q}_j}(s) & = & 
- \frac{4 \pi \alpha_{em} e_{q}}{s} \; + \;
  \frac{4 \pi \alpha_{em}}{sin^{2}\theta_{W} \cdot cos^{2}\theta_{W} } 
\cdot \frac{g^{e}_{i} g^{q}_{j}}
{s - M_{Z}^{2} + i s \frac{\Gamma_{Z}}{M_{Z}}}
  \; + \; \eta^{eq}_{ij} \label{eq-ms}
\end{eqnarray}
where $s$ is the center-of-mass energy squared of the four-fermion reaction.
The sign of the contact interaction contribution to the $s$-channel 
amplitude  \eqref{eq-ms} is the same as for the $t$-channel amplitude
\eqref{eq-mt}. However, the Standard Model amplitude changes its sign 
due to the opposite signs of $s$ and $t$ variables.
It is therefore important to notice that
the resulting sign of the interference terms in the cross-section
for $e^{\pm}p$ scattering is different from that in $e^{+}e^{-}$ or $p\bar{p}$
scattering.

The contact interaction coupling strength $\eta$ can be related to 
the mass scale\footnote{Exchanged particle mass or compositeness scale.} 
${\cal M}$ of new physics through the formula:
\begin{eqnarray}
\eta & = & \pm \frac{g_{CI}^{2}}{{\cal M}^{2}} \nonumber
\end{eqnarray}
where $g_{CI}$ is the unknown coupling strength of new interactions. 
As the contact interaction contribution always depends on the $g_{CI}$ 
to ${\cal M}$ ratio, it is convenient to consider the effective 
mass scale $\Lambda$ defined through the formula:
\begin{eqnarray}
\eta & = & \pm \frac{4 \pi}{\Lambda^{2}}  \nonumber
\end{eqnarray}
which corresponds to the choice $g^{2}_{CI}=4\pi$.

\subsection{General Model}

In the most general case, vector contact interactions are described by
4 independent couplings for every lepton-quark pair. 
With only 2 lepton ($e$ and $\mu$) and 
5 quark flavours (i.e. neglecting $t$ quark contribution),
we still have 40 independent couplings.
It would be very difficult, if not impossible, to consider 
the model with 40 free parameters.
However, some of these parameters (couplings) are weakly constrained 
by existing experimental data.
To reduce the number of the free model parameters, weakly constrained
couplings can be either neglected  
or additionally constrained by relating them to some other couplings.

Most of existing  experimental data is sensitive predominantly 
to electron-up and  electron-down quark couplings. 
Therefore, the first model
considered in this analysis is the one assuming that these 8 couplings 
($\eta^{ed}_{LL}$, $\eta^{ed}_{LR}$, $\eta^{ed}_{RL}$, $\eta^{ed}_{RR}$,
$\eta^{eu}_{LL}$, $\eta^{eu}_{LR}$, $\eta^{eu}_{RL}$, $\eta^{eu}_{RR}$)
can vary independently, whereas other couplings (for $s$, $c$, $b$, $t$ 
quarks and/or $\mu$, $\tau$ leptons) are assumed to vanish.
This case will be referred to as {\bf the general model}.

The other possibility is to impose additional relations between couplings.
The common choice is to assume lepton universality:
\begin{eqnarray}
\eta^{eq}_{ij} & = \eta^{\mu q}_{ij} & = \eta^{\tau q}_{ij} \label{eq-lu}
\end{eqnarray}
and quark family universality:
\begin{eqnarray}
\eta^{eu}_{ij} & = \eta^{ec}_{ij} & = \eta^{et}_{ij} \label{eq-qu} \\
\eta^{ed}_{ij} & = \eta^{es}_{ij} & = \eta^{eb}_{ij} \nonumber
\end{eqnarray}
Lepton universality allows us to include data on muon pair production at
the Tevatron (see section \ref{sec-dy}), whereas assuming quark family 
universality significantly improves the constraints which we can obtain from
LEP2 measurements (see section \ref{sec-lep}).
As a result, experimental constraints on contact interactions can be 
significantly improved without increasing the number of free model parameters.
The model assuming relations \eqref{eq-lu} and \eqref{eq-qu} will be 
referred to as {\bf the model with family universality}.

\subsection{$SU(2)_{L} \times U(1)_{Y}$ Universality}
\label{sec-su2}

Another commonly used assumption about lepton-quark contact interactions 
is that they satisfy the $SU(2)_{L} \times U(1)_{Y}$  gauge invariance
of the Standard Model. Assuming that left-handed electrons and quarks 
belong to $SU(2)_{L}$ doublets and that the contact interaction 
Lagrangian \eqref{eq-lagr} respects the $SU(2)_{L}$ symmetry implies 
a relation between contact 
terms involving left-handed $u$ and $d$ quarks \cite{ciglob}: 
\begin{eqnarray}
\eta^{eu}_{RL} & = & \eta^{ed}_{RL} \nonumber 
\end{eqnarray}
which reduces the number of free model parameters from 8 to 7.
The $SU(2)_{L} \times U(1)_{Y}$ also  relates $eeqq$ contact interaction
couplings with those of $\nu \nu qq$ interactions
\begin{eqnarray}
\eta^{\nu u}_{LL} & = & \eta^{ed}_{LL} \label{eq-etanu}  \\
\eta^{\nu d}_{LL} & = & \eta^{eu}_{LL} \nonumber  \\
\eta^{\nu u}_{LR} & = & \eta^{eu}_{LR} \nonumber  \\
\eta^{\nu d}_{LR} & = & \eta^{ed}_{LR} \nonumber 
\end{eqnarray}
This allows us to use, in the study of $eeqq$ contact interactions,
additional data on NC neutrino scattering (see section \ref{sec-le}).

Moreover, assuming the $SU(2)_{L} \times U(1)_{Y}$  universality 
introduces a related contact interaction term in the Charged
Current process $eq \rightarrow \nu q'$. The coupling constant for
the induced Charged Current contact interaction is
\begin{eqnarray}
\eta^{CC} \equiv \eta^{eu \nu d} & = & 
              \eta^{ed}_{LL} - \eta^{eu}_{LL} \label{eq-cc} 
\end{eqnarray}
This relation allows us to use, in the study of Neutral Current 
contact interaction, also data from Charged Current processes
(see section \ref{sec-hera} and \ref{sec-le}).

The model assuming the $SU(2)_{L} \times U(1)_{Y}$  universality will 
be referred to as {\bf the SU(2) model}.
In order to reduce the number of models,
the SU(2) models considered in this analysis always assume lepton
and quark family universality.

\subsection{One-parameter models}

Using data from a single experiment it is mostly not possible to put 
significant constraints on contact interaction scales in the general 
case. Therefore it is a common practise to consider particular models, 
which assume fixed relations between the separate couplings,
reducing the number of free parameters to one. 
For example, the so called vector-vector model assumes that all couplings 
are equal:
\begin{eqnarray}
\eta^{ed}_{LL} = \eta^{ed}_{LR} = \eta^{ed}_{RL} = \eta^{ed}_{RR} =
\eta^{eu}_{LL} = \eta^{eu}_{LR} = \eta^{eu}_{RL} = \eta^{eu}_{RR}
                    & \equiv & \eta_{VV}  \nonumber
\end{eqnarray}
Mass scale limits obtained in one-parameter models are, artificially,
much stronger than in the general model. They will be considered in this
analysis to allow comparison with other results. The relations between
couplings assumed for different models are listed in Table 
\ref{tab-models}\cite{zeusci}. 
It should be noticed that all one-parameter models 
considered assume
\begin{eqnarray}
\eta^{eq}_{LL} + \eta^{eq}_{LR} - \eta^{eq}_{RL} - \eta^{eq}_{RR}
                    & = &  0  \nonumber
\end{eqnarray}
for $q=u,d$, to avoid strong limits coming from atomic parity 
violation measurements (see section \ref{sec-le}).
For all one-parameter models quark and lepton family universality 
is assumed. 
The results obtained both with and without  imposing 
the $SU(2)_{L} \times U(1)_{Y}$ universality  are presented, except for
the U2, U4 and U6 models, which violate it explicitly 
($\eta^{eu}_{RL} \ne  \eta^{ed}_{RL}$). 

\begin{table}[tp]
  \begin{center}
   \begin{tabular}{crrrrrrrr}
      \hline\hline\hline\noalign{\smallskip}
  Model & 
$\eta^{ed}_{LL}$ & $\eta^{ed}_{LR}$ & $\eta^{ed}_{RL}$ & $\eta^{ed}_{RR}$ & 
$\eta^{eu}_{LL}$ & $\eta^{eu}_{LR}$ & $\eta^{eu}_{RL}$ & $\eta^{eu}_{RR}$ \\ 
\hline\hline\hline\noalign{\smallskip}
VV & $+\eta$& $+\eta$& $+\eta$& $+\eta$& $+\eta$& $+\eta$& $+\eta$& $+\eta$\\
AA & $+\eta$& $-\eta$& $-\eta$& $+\eta$& $+\eta$& $-\eta$& $-\eta$& $+\eta$\\
VA & $+\eta$& $-\eta$& $+\eta$& $-\eta$& $+\eta$& $-\eta$& $+\eta$& $-\eta$\\
\hline\noalign{\smallskip}
X1 & $+\eta$ & $-\eta$ &  &  & $+\eta$ & $-\eta$ &  &  \\
X2 & $+\eta$ &  & $+\eta$ &  & $+\eta$ &  & $+\eta$ &  \\
X3 & $+\eta$ &  &  & $+\eta$ & $+\eta$ &  &  & $+\eta$ \\
X4 &  & $+\eta$ & $+\eta$ &  &  & $+\eta$ & $+\eta$ &  \\
X5 &  & $+\eta$ &  & $+\eta$ &  & $+\eta$ &  & $+\eta$ \\
X6 &  &  & $+\eta$ & $-\eta$ &  &  & $+\eta$ & $-\eta$ \\ 
\hline\noalign{\smallskip}
U1 &   &   &   &   & $+\eta$ & $-\eta$ &  &  \\
U2 &   &   &   &   & $+\eta$ &  & $+\eta$ &  \\
U3 &   &   &   &   & $+\eta$ &  &  & $+\eta$ \\
U4 &   &   &   &   &  & $+\eta$ & $+\eta$ &  \\
U5 &   &   &   &   &  & $+\eta$ &  & $+\eta$ \\
U6 &   &   &   &   &  &  & $+\eta$ & $-\eta$ \\ 
\hline\hline\hline\noalign{\smallskip}
    \end{tabular}
  \end{center}
  \caption{Relations between couplings for the one-parameter models
           considered in this paper.}
  \label{tab-models}
\end{table}

%---------------------------------------------------------------------------

\section{Experimental Data}
\label{sec-data}

In this section the data used to constrain contact interaction model
are presented. 
For each measurement, the formula describing the possible influence 
of the new couplings on the measured quantities is given.
Description of the statistical methods used to interpret the data
will be presented in the section \ref{sec-method}.

\subsection{High-$Q^{2}$ DIS at HERA}
\label{sec-hera}

Used in this analysis are the latest data on high-$Q^{2}$ $e^{+}p$ NC DIS
from both H1 \cite{h1_98} and ZEUS \cite{zeus98}, corresponding to 
integrated data luminosities of 37 and 47$pb^{-1}$, respectively. 
Older results from $e^{-}p$ NC DIS scattering \cite{h1ep,zeusep}
are also used, although the influence of these data is marginal. 
For models with the $SU(2)_{L} \times U(1)_{Y}$  universality, 
as mentioned in section \ref{sec-model},
data on $e^{+}p$ CC DIS \cite{h1_98,zeuscc} are also included in the fit.

HERA experiments quote their high-$Q^{2}$ DIS results in terms of
numbers of events and/or cross-sections\footnote{If not given,
the number of events can be estimated from the cross-section value assuming 
that the statistical error quoted corresponds to the Poisson error 
on the number of measured events, $\sigma_{N} = \sqrt{N}$.}
 measured in bins of $Q^{2}$.
For simplicity let us consider a single $Q^{2}$ bin ranging from
$Q^{2}_{min}$ to $Q^{2}_{max}$. Assume that  $n_{SM}$ events are
expected from the Standard Model. 

The leading order doubly-differential cross-section for positron-proton 
NC DIS ($e^{+}p \rightarrow e^{+} X$) can be written as \cite{ciglob}
\begin{eqnarray}
\frac{d^{2}\sigma^{LO}}{dx dQ^{2}} & = &
\frac{1}{16\pi} \sum_{q} 
q(x) \left\{ |M^{eq}_{LR}|^{2} + |M^{eq}_{RL}|^{2} + 
(1-y)^{2} \left[  |M^{eq}_{LL}|^{2} + |M^{eq}_{RR}|^{2} \right] \right\}
\; +  \nonumber \\
 & &  ~~~~~~~~~~\bar{q}(x)\left\{ |M^{eq}_{LL}|^{2} + |M^{eq}_{RR}|^{2} + 
(1-y)^{2} \left[  |M^{eq}_{LR}|^{2} + |M^{eq}_{RL}|^{2} \right] \right\}
 \nonumber
\end{eqnarray}
where $x$ is the Bjorken variable, describing the fraction of proton momentum 
carried by a quark (antiquark), $q(x)$ and $\bar{q}(x)$ are
the quark and antiquark momentum distribution functions in the proton and
$M^{eq}_{ij}$ are the scattering amplitudes of
equation \eqref{eq-mt}, which can include contributions from contact
interactions described by a set of couplings $\vec{\eta}$.

The cross-section (including the contribution from contact interactions),
integrated over the $x$ and $Q^{2}$ range of an experimental $Q^{2}$ bin is
\begin{eqnarray}
\sigma^{LO} ( \vec{\eta} ) & = & 
\bigint{Q^{2}_{min}}{Q^{2}_{max}} dQ^{2} 
\bigint{\frac{Q^{2}}{s \cdot y_{max}}}{1} dx \;  
         \frac{d^{2}\sigma^{LO} ( \vec{\eta}) }{dx dQ^{2}} \label{eq-intdis}
\end{eqnarray}
where $y_{max}$ is the upper limit on reconstructed Bjorken variable $y$
imposed in the analysis\footnote{For the NC DIS analysis H1 uses $y_{max}=0.9$,
whereas ZEUS uses $y_{max}=0.95$. For CC DIS analysis both experiments use 
$y_{max}=0.9$.}.
The number of events expected from the Standard Model with contact 
interaction contributions can now be calculated as:
\begin{eqnarray}
n(\vec{\eta}) & = & n_{SM} \cdot 
\left( \frac{\sigma^{LO}(\vec{\eta})}{\sigma^{LO}_{SM}} \right)  
                                  \label{eq-neta}
\end{eqnarray}
where $\sigma^{LO}_{SM}$ is the Standard Model cross-section
calculated with formula \eqref{eq-intdis} (setting $\vec{\eta}=\vec{0}$).
Leading-order expectations of the contact interaction models 
are used to rescale the Standard Model prediction $n_{SM}$ 
coming from detailed experiment simulation. 
This accounts not only for different experimental effects, but also for
higher order QCD and electroweak corrections.
Validity of this approach is discussed in section \ref{sec-method}.

\subsection{Drell-Yan lepton pair production at the Tevatron}
\label{sec-dy}

Used in this analysis are data on Drell-Yan lepton pair 
production from the CDF \cite{dy_cdf} and D0 \cite{dy_d0} experiments.
Both experiments present numbers of measured high-mass electron
pairs ($p \bar{p} \rightarrow e^{+} e^{-}\;X$). CDF also presents
results on muon pair production 
($p \bar{p} \rightarrow \mu^{+} \mu^{-}\;X$), which are used in the case
of models with family universality (see section \ref{sec-model}).

The leading order cross-section for lepton pair production in 
$p \bar{p}$ collisions is
\begin{eqnarray}
\frac{d^{2}\sigma^{LO}}{dM_{ll} dY} & = &
\frac{M_{ll}^{3}}{72\pi s} \sum_{q} q(x_{1}) q(x_{2}) 
      \sum_{i,j=L,R} |M^{eq}_{i j}|^{2}  \nonumber
\end{eqnarray}
where $M_{ll}$ is the invariant mass of lepton pair, $Y$ is the rapidity 
of the lepton pair center-of-mass frame, $x_{1}$ and $x_{2}$ are
the fractions of proton and antiproton momenta carried by the annihilating
quarks. The scattering amplitudes 
$M^{eq}_{i j}$ and the parton density functions are calculated for scale
\begin{eqnarray}
\hat{s} & = & x_{1} x_{2} s  \nonumber
\end{eqnarray}
where $s$ is the total proton-antiproton center of mass energy squared.

The cross-section corresponding to the $M_{ll}$ range from $M_{min}$ to
$M_{max}$ is calculated as
\begin{eqnarray}
\sigma^{LO} ( \vec{\eta} ) & = & 
\bigint{M_{min}}{M_{max}} dM_{ll} 
\bigint{-Y_{max}}{Y_{max}} dY  \; A_{ll}(Y) \cdot
         \frac{d^{2}\sigma ( \vec{\eta}) }{dM_{ll} dY} \label{eq-intdy}
\end{eqnarray}
where $Y_{max}$ is the upper limit on the rapidity of the produced 
lepton pair:
\begin{eqnarray}
Y_{max} & = & \ln \frac{\sqrt{s}}{M_{ll}} \; , \nonumber 
\end{eqnarray}
and $A_{ll}(Y)$ is the acceptance function, resulting from the integration
over the lepton pair production angle in the center of mass system,
with angular detector coverage taken into account.
The cross-section calculated with equation \eqref{eq-intdy} is used to 
calculate the number of events expected from the Standard Model with contact 
interaction contributions using formula \eqref{eq-neta}.

\subsection{Measurements from LEP}
\label{sec-lep}

Many measurements at LEP  are sensitive to different kinds of "new physics".  
The $eeqq$ contact interactions can be directly tested in the
measurement of the total cross-section
for $e^{+} e^{-} \rightarrow q \bar{q}$. 
Using flavour tagging techniques, additional constraints can be
obtained from the measurement of the heavy quark decay fractions $R_{b}$
and $R_{c}$, and of the forward-backward asymmetries $A^{q}_{FB}$ 
of $q\bar{q}$ events.

The leading order formula for the total quark pair production
cross-section  $e^{+} e^{-} \rightarrow q \bar{q}$, at the total
electron-positron center of mass energy squared $s$, is
\begin{eqnarray}
\sigma^{LO}(s) & = &
\frac{s}{16\pi} \sum_{q} \sum_{i,j=L,R} 
|M^{eq}_{i j}|^{2}                         \label{eq-lep}
\end{eqnarray}
where $M^{eq}_{ij}$ are 
the scattering amplitudes described by equation \eqref{eq-ms}, 
including contributions from contact interaction couplings $\vec{\eta}$.
For comparison with measured experimental values, the leading order
contact interaction cross-sections are rescaled using the expected 
Standard Model cross-section $\sigma^{SM}(s)$ quoted by experiments:
\begin{eqnarray}
\sigma(s,\vec{\eta}) & = &
\sigma^{SM}(s) \cdot \left(
\frac{\sigma^{LO}(s,\vec{\eta})}{\sigma^{LO}(s,0)} \right) \label{eq-seta}
\end{eqnarray}
where $\sigma^{LO}(s,0)$ is the leading-order Standard Model 
cross-section ($\vec{\eta}=\vec{0}$), calculated with
equation \eqref{eq-lep}.
This takes into account possible experimental effects and
higher order QCD and electroweak corrections 
(for discussion see section \ref{sec-method}).
All four LEP experiments
have recently presented data on $\sigma_{had}$ for center-of-mass
energies up to 189~GeV \cite{aleph,delphi,delphi2,l3,opal}.

The sensitivity of the total hadronic cross-section to the contact
interaction coupling strength $\vec{\eta}$ is limited by the fact that
the interference terms in the quark-pair production cross-sections have
opposite signs for up-type and down-type quarks.
In the total cross-section, summed over all quark 
flavours\footnote{Production of the $t$ quark is taken into account
only for $\sqrt{s} >$ 350 GeV.}, these terms
tend to compensate each other.
However, if this is the case, the fraction of events produced with the 
given quark-pair flavour turns out to be very sensitive to the contact 
interaction couplings.

Using different flavour tagging techniques, cross-sec\-tions for
$b\bar{b}$  and $c\bar{c}$ pair production and the corresponding
fractions $R_{b}$ and $R_{c}$ can be measured.
Although the limited tagging efficiency and purity significantly
affects the measurement, useful constraints on contact interaction
couplings can be extracted. 
Used in this analysis are results on $R_{b}$ coming from 
{\sc Aleph}\cite{aleph,alephrb}, {\sc Delphi} \cite{delphi} and 
{\sc Opal} \cite{opalrb} as well as {\sc Delphi} results on 
$R_{c}$ \cite{delphi}.

The contact interaction contribution to the scattering amplitude affects
also the observed forward-backward asymmetry of $q\bar{q}$ events.
In the leading order the forward-backward asymmetry can be calculated
as
\begin{eqnarray}
A^{q}_{FB}(s) & = & \frac{3}{4} \cdot \frac{
|M^{eq}_{LL}|^{2} - |M^{eq}_{LR}|^{2} - |M^{eq}_{RL}|^{2} + |M^{eq}_{RR}|^{2}
}{
|M^{eq}_{LL}|^{2} + |M^{eq}_{LR}|^{2} + |M^{eq}_{RL}|^{2} + |M^{eq}_{RR}|^{2}
} \nonumber
% ~~~~~~ \label{eq-afb}
\end{eqnarray}
where the factor $\frac{3}{4}$ corresponds to the integration over the
full angular range\footnote{For results which are based on the sample of
events selected with $| \cos \theta | < 0.9$, this factor is reduced to
0.70866... }.

Constraints upon the forward-backward  asymmetries $A^{q}_{FB}$ are
obtained using a jet charge technique.
After clustering all the events into two jets, the jet charge $Q_{jet}$
of each jet can be determined from the momentum weighted sum over all
charged tracks in the jet.
The sign of $Q_{jet}$ coincides with the charge
of produced quark in about 70\% of events. 
The forward-backward asymmetry for the selected sample of events
(e.g. $b$-tagged events) can be extracted in two ways.
The method used by {\sc Aleph} is based on the measurement of the mean
charge difference between the forward and backward jets $\langle Q_{FB}
\rangle = \langle Q^{F}_{jet} \rangle  - \langle Q^{B}_{jet} \rangle $.
{\sc Delphi} and {\sc Opal} extract $A^{q}_{FB}$ from the angular 
distribution of jets with well defined sign.
In both cases, the measured asymmetry depends on the parton-level
asymmetries $A^{q}_{FB}$ and on the quark content of the selected
sample.
As the up-type and down-type quarks have charges of opposite signs,
the measured asymmetry is very sensitive to the relative contribution
of different quark flavours.
Even if we measure the asymmetry for the flavour-tagged sample,
the selected sample of events is always contaminated by other quark
flavours (e.g. a $b$-tagged sample always contains a fraction of $c\bar{c}$
events) and the measured value depends strongly on the quark production
fractions (e.g. $R_{b}$ and $R_{c}$).
This is the reason why the measurement of the forward-backward
asymmetry is very sensitive to the contact interaction couplings.
Used in this analysis are the measurements of forward-backward
asymmetry for the $b$-tagged events \cite{aleph,delphi,opalrb},
$c$-tagged events \cite{delphi} and anti-tagged events
\cite{aleph,delphi}. 

\subsection{Data from low energy experiments}
\label{sec-le}

The low energy data are included in the present analysis in the manner
which follows closely the approach presented in \cite{ciglob,cile}.
Therefore only basic assumptions are listed here 
and technical details are omitted.

In case of the general contact interaction model the following
constraints from low energy experiments are considered:
\begin{itemize}
\item Atomic Parity Violation (APV) \\
The Standard Model predicts parity non-conservation in atoms caused
(in lowest order) by the $Z^{\circ}$ exchange between
electrons and quarks in the nucleus. 
Experimental results on parity violation in atoms are given in terms
of the weak charge $Q_{W}$ of the nuclei. A very precise determination
of $Q_{W}$ for Cesium atoms was recently reported \cite{apvcs}.
The experimental result differs from the Standard Model 
prediction \cite{apvth,rpp} by:
\begin{eqnarray}
\Delta Q_{W}^{Cs}  \equiv Q_{W}^{meas} - Q_{W}^{SM} 
                      & = & 0.71 \pm 0.84    \nonumber
\end{eqnarray}
Corresponding results have also been obtained for thallium \cite{apvtl,rpp}:
\begin{eqnarray}
\Delta Q_{W}^{Tl}  & = & 1.9 \pm 3.6 \; .   \nonumber
\end{eqnarray}
These measurements are used to place limits on 
contact interaction contributions to $Q_{W}$:
\begin{eqnarray}
\Delta Q_{W}(\vec{\eta}) & = & 
\frac{2Z+N}{\sqrt{2} G_{F}}
 \left(\eta^{eu}_{LL}+\eta^{eu}_{LR}-\eta^{eu}_{RL}-\eta^{eu}_{RR} \right)
                                   \nonumber \\
 & + &
\frac{Z+2N}{\sqrt{2} G_{F}}
 \left(\eta^{ed}_{LL}+\eta^{ed}_{LR}-\eta^{ed}_{RL}-\eta^{ed}_{RR} \right)
                 \nonumber
\end{eqnarray}

\item electron-nucleus scattering \\
The limits on possible contact interaction contributions to 
electron-nucleus scattering at low energies can be extracted from
the polarisation asymmetry measurement
\begin{eqnarray}
A  & = & 
\frac{d\sigma_{R}-d\sigma_{L}}{d\sigma_{R}+d\sigma_{L}} \nonumber
\end{eqnarray}
where $d\sigma_{L(R)}$ denotes the differential cross-section of
left- (right-) handed electron scattering. 
Polarisation asymmetry directly measures the parity violation
resulting from the interference between the weak ($Z^{\circ}$ exchange)
and the electro-magnetic ($\gamma$ exchange) scattering amplitudes.
For isoscalar targets, taking into account valence quark contributions only,
the polarisation asymmetry for elastic electron scattering is
\begin{eqnarray}
A_{el}  & = & 
-\frac{3 \sqrt{2} G_{F} Q^{2}}{20 \pi \alpha_{em} } 
\left[ 2 \left( g^{u}_{L} + g^{u}_{R} \right) - 
         \left( g^{d}_{L} + g^{d}_{R} \right) \right]  \nonumber
\end{eqnarray}
where $Q^{2}$ is the four-momentum transfer and $g^{q}_{i}$ are 
quark electroweak couplings, as introduced in equation \eqref{eq-gdef}.
Contact interactions modify the effective quark electroweak coupling
\begin{eqnarray}
 \bigmod{g^{q}_{i}}{ef\!f} & = & 
  g^{q}_{i} - \frac{\eta^{e q}_{Li}}{2\sqrt{2} G_{F}} \label{eq-geff}
\end{eqnarray}
The constraints used in this analysis come from the SLAC $e$D 
experiment \cite{slac}, the Bates $e$C experiment \cite{bates} and 
the Mainz experiment on $e$Be scattering \cite{mainz}. 
In case of models with family universality 
also data from the  $\mu^{\pm}$C experiment at CERN\cite{cern} are 
included\footnote{The constraints from the $\mu^{\pm}$C experiment 
result from the comparison of $\mu^{+}_{L} N$ and $\mu^{-}_{R} N$ cross
sections.}.

\end{itemize}

In case of the SU(2) models additional constraints come from:
\begin{itemize}

\item neutrino-nucleus scattering \\
Constraints on the couplings of quarks to the $Z^{\circ}$ 
and/ or additional $\nu \nu qq$ contact interactions (related to 
$eeqq$ CI, as described in section \ref{sec-su2}) can also be
derived from the precise measurement of the ratio of Neutral Current
to Charged Current neutrino-nucleon scattering cross sections
\begin{eqnarray}
R^{\nu} & = & \frac{\sigma^{\nu N}_{_{NC}}}{\sigma^{\nu N}_{_{CC}}} \; .
          \nonumber
\end{eqnarray}
However, when using constraints on $g^{q}_{i}$ resulting from measurement
of $R^{\nu}$, one also has to take into account that possible 
contact interaction contribution affects not only the Neutral Current 
but also the Charged Current scattering cross-section 
(see section \ref{sec-su2}).
Therefore the quark electroweak coupling extracted from $R^{\nu}$
measurements should be expressed as\footnote{This correction 
seems to be missing in \cite{ciglob,cile}.}
\begin{eqnarray}
 \bigmod{g^{q}_{i}}{meas} & = & 
  \frac{g^{q}_{i} - \frac{\eta^{\nu q}_{Li}}{2\sqrt{2} G_{F}}}
        {1 - \frac{\eta^{CC}}{4\sqrt{2} G_{F}}} \nonumber
\end{eqnarray}
It is important to notice that $\eta^{e q}_{Li}$ entering 
formula \eqref{eq-geff} has been replaced here by $\eta^{\nu q}_{Li}$.
This is because the effective $g^{u}_{L}$ and $g^{d}_{L}$ couplings
measured in neutrino scattering are sensitive to ``flavour crossed''
contact interaction couplings $\eta^{ed}_{LL}$ and $\eta^{eu}_{LL}$ 
respectively, which results from relations \eqref{eq-etanu}.
Experimental constraints on $R^{\nu}$ come mainly from
muon-neutrino experiments.
Assuming lepton and quark family universality,
the following measurements of $g^{q}_{i}$ from $\nu_{\mu} N$ 
scattering are used: 
the results compiled by Fogi and  Haidt \cite{fh}
and the recent constraints from CCFR\cite{ccfr} and NuTeV\cite{nutev}.

\item lepton-hadron universality of weak Charged Currents \\
Charged Current contact interactions which are 
induced by  $SU(2)_{L} \times U(1)_{Y}$  
universality (see equation \eqref{eq-cc}) would also affect the
measurement of $V_{ud}$ element of the Cabibbo-Kobayashi-Maskawa (CKM) 
matrix, leading to the effective violation of unitarity
\cite{cickm,cickm2}. The current experimental constraint is \cite{rpp}
\begin{eqnarray}
|V_{ud}|^{2} + |V_{us}|^{2} + |V_{ub}|^{2} & = & 0.9969 \pm 0.0022 \nonumber
\end{eqnarray}
whereas the expected contribution from the contact interaction is
\begin{eqnarray}
\bigmod{V_{ud}}{meas} & = & 
V_{ud}^{SM} \cdot \left( 1 - \frac{\eta^{CC}}{4\sqrt{2}G_{F}} \right) 
\nonumber
\end{eqnarray}

\item electron-muon universality \\
In the similar way Charged Current contact interactions
would also lead to effective violation of $e$-$\mu$ universality
in charged pion decay \cite{cickm}. The current experimental value of
$R=\Gamma(\pi^{-} \rightarrow e \bar{\nu})/
   \Gamma(\pi^{-} \rightarrow \mu \bar{\nu})$ is \cite{emu}
\begin{eqnarray}
\frac{R_{meas}}{R_{SM}} & = & 0.9966 \pm 0.030 \nonumber
\end{eqnarray}
whereas the expected contribution from the contact interaction is
\begin{eqnarray}
\bigmod{R}{meas} & = & 
R_{SM} \cdot \left( 1 - \frac{\eta^{CC}}{4\sqrt{2}G_{F}} \right)^{2} 
\nonumber
\end{eqnarray}

\end{itemize}

It is interesting to notice, that data in Charged Current sector 
may point to a slight violations of the unitarity of the CKM matrix 
and of the $e$-$\mu$ universality. Both measurements are
consistent with the presence of CC contact interactions with a mass 
scale of the order of 10~TeV. 
The combined significance of these two results is about 1.8$\sigma$,
but it has a considerable influence on global analysis results 
for the SU(2) model.

%---------------------------------------------------------------------------

\section{Analysis method}
\label{sec-method}

\rightmark

The aim of this study is to find the allowed ranges for contact interaction
couplings within the different models considered. To do so,
the probability function in the coupling space,
\begin{eqnarray}
{\cal P}(\vec{\eta}) & \sim & \prod_{i} P_{i}(\vec{\eta}), \label{eq-prob}
\end{eqnarray}
is calculated. In \eqref{eq-prob}, the product runs over all experimental 
data $i$ and $\vec{\eta}$ represents the set of free parameters for 
a given model (one or many).
This section describes how the probability function is defined and which
corrections are included to take into account experimental conditions.

\subsection{Statistical errors}
\label{sec-met-stat}

All experimental data used in this analysis can be divided into two
classes.

\begin{enumerate}
\item
For experiments in which a result can be presented as 
a single number with an error which is considered to reflect a Gaussian
probability distribution, the constraints on the contact interaction couplings 
can be usually expressed using the equation
\begin{eqnarray}
F(\vec{\eta}) & =  & \Delta A \; \pm \; \sigma_{A} \nonumber
\end{eqnarray}
where $\Delta A$ is the difference between the measured value and the Standard
Model prediction, and $F(\vec{\eta})$ is the expected contact interaction 
contribution to the measured value of A. The resulting probability function 
can be written as
\begin{eqnarray}
P_{i}(\vec{\eta}) & \sim & 
exp \left( -\frac{1}{2} 
\frac{(F(\vec{\eta}) - \Delta A)^{2}}{\sigma_{A}^{2}} \right)
\label{eq-gauss}
\end{eqnarray}
reflecting the definition of the Gaussian error $\sigma_{A}$.
This approach is used for all low energy data as well as for
the LEP hadronic cross-section measurements.

\item
On the other hand, when the experimentally measured quantity is 
the number of events of a particular kind (e.g. HERA high-$Q^{2}$ events
or Drell-Yan lepton pairs at the Tevatron), and especially when this number
is small, the probability is better described by the Poisson
distribution
\begin{eqnarray}
P_{i}(\vec{\eta}) & \sim & 
\frac{ n(\vec{\eta})^{N} \cdot exp( -n(\vec{\eta})) }{ N ! }
\label{eq-poisson}
\end{eqnarray}
where $N$ and $n(\vec{\eta})$ are the measured and expected 
number of events in a given experiment, respectively, and $n(\vec{\eta})$
takes into account a possible
contact interaction contribution. This approach has been used for
HERA and the Tevatron data.

\end{enumerate}

\subsection{Systematic errors}

For low energy data the total measurement error can be used in 
formula \eqref{eq-gauss} taking into account both statistical and
systematic errors.
For collider data, formula \eqref{eq-gauss} or \eqref{eq-poisson} 
is used to take into account the statistical error of the measurement only. 
As for the systematic errors, it is assumed
that within a given data set (e.g. $e^{+}p$ NC DIS data from ZEUS ) they
are correlated to 100\%.
This seems to be a much better approximation
of the experimental conditions than assuming that systematic errors
are uncorrelated\footnote{Unfortunately the experiments do not publish 
the correlation matrix for their systematic errors so these are the 
only possible choices.}. 
In fact most of the contributing systematic 
uncertainties at HERA are highly correlated between different $Q^{2}$
bins, as for example energy scale uncertainty or the 
luminosity measurement.

For each data set, a common systematic shift parameter $\delta$ has been
introduced to describe the possible variation of event numbers expected 
at HERA or the Tevatron, or cross-sections predicted at LEP,
due to systematic error:
\begin{eqnarray}
n_{SM} & = & \bar{n}_{SM} \; + \; \delta \cdot \sigma_{n}^{sys} \nonumber \\
{\rm or }\;\;\;\;\;\;\;\;\;
\sigma^{SM} & = & \bar{\sigma}_{SM} \; + \; 
                          \delta \cdot \sigma_{\sigma}^{sys} \nonumber 
\end{eqnarray}
where $\bar{n}_{SM}$ ($\bar{\sigma}_{SM}$) is the nominal expectation 
from the Standard Model and $\sigma_{n}^{sys}$ ($\sigma_{\sigma}^{sys}$) 
is the total systematic uncertainty attributed to this number.
Parameters $\delta$ can been treated as additional 
free parameters when maximising the overall model probability 
${\cal P}(\vec{\eta})$.
When doing so, normal probability distributions for parameters 
$\delta$ are included in the definition \eqref{eq-prob} 
of the probability function\footnote{This corresponds to 
the assumption that systematic errors are described by the Gaussian
probability distribution}.

\subsection{Migration corrections}

Equation \eqref{eq-neta} introduced in section \ref{sec-data}
takes into account experimental conditions at HERA and the Tevatron.
The number of events expected with contact interaction contribution is
calculated by rescaling the Standard Model prediction $n_{SM}$ 
coming from  the detailed experiment simulation. 
However, this is only an approximation based on the assumption
that the acceptance for contact interaction events 
is the same as for standard NC DIS events. 
Although the detection efficiency for given ($x$,$Q^{2}$) or 
($M_{ll}$,$Y$) is always the same (as we have the same final state), 
the distribution of events in the kinematic plane in the presence 
of the contact interactions can differ significantly.  
This can affect the measurement due to the finite $Q^{2}$ 
or $M_{ll}$ resolution. 
To take this effect into account a dedicated migration correction is
introduced.
The DIS cross-section in  the $Q^{2}$ bin 
from $Q^{2}_{min}$ to $Q^{2}_{max}$ is calculated by the following
extension of eq. \eqref{eq-intdis}:
\begin{eqnarray}
\sigma^{LO}_{DIS} ( \vec{\eta} ) & = & 
\bigint{0}{s} dQ^{2} \cdot S(Q^{2};Q^{2}_{min},Q^{2}_{max},\sigma_{Q^{2}}) \;
\bigint{\frac{Q^{2}}{s \cdot y_{max}}}{1} dx  
\frac{d^{2}\sigma ( \vec{\eta}) }{dx dQ^{2}} 
\nonumber
\end{eqnarray}
where $\sigma_{Q^{2}}$ is the $Q^{2}$ resolution, as quoted by
experiments, assumed to be constant within the bin. 
The Drell-Yan cross-section is calculated by the similar
extension of eq. \eqref{eq-intdy}:
\begin{eqnarray}
\sigma^{LO}_{DY} ( \vec{\eta} ) & = & 
\bigint{0}{\sqrt{s}} dM_{ll} 
\cdot S(M_{ll};M_{min},M_{max},\sigma_{M}) \; 
         \bigint{-Y_{max}}{Y_{max}} dY  \; A_{ll}(Y) \cdot
         \frac{d^{2}\sigma ( \vec{\eta}) }{dM_{ll} dY} 
\nonumber
\end{eqnarray}
where $\sigma_{M}$ is the $M_{ll}$ resolution.
The mass resolution has been estimated from the quoted
calorimeter energy resolution (for electrons) or tracking momentum
resolution (for muon pairs).
The acceptance function used in both formula
\begin{eqnarray}
S(x;a,b,\sigma) & = & \bigint{-\infty}{x} dy \;
\frac{1}{\sqrt{2 \pi} \sigma } \left[
exp \left( -\frac{1}{2} 
\frac{(y-a)^{2}}{\sigma^{2}} \right) \; - \;
exp \left( -\frac{1}{2} 
\frac{(y-b)^{2}}{\sigma^{2}} \right)
\right]  \nonumber
\end{eqnarray}
describes the probability that the true value $x$ measured with 
resolution $\sigma$ will be reconstructed between $a$ and $b$.
The migration corrections are important for the muon-pair production
results from the Tevatron and for the CC DIS results from HERA.
For electron-pair production or for NC DIS results, when the
corresponding mass and Q$^{2}$ resolutions are much better, the effects
of the migration corrections are very small. 

\begin{figure}[tp]
\centerline{\resizebox{\figwidth}{!}{%
  \includegraphics{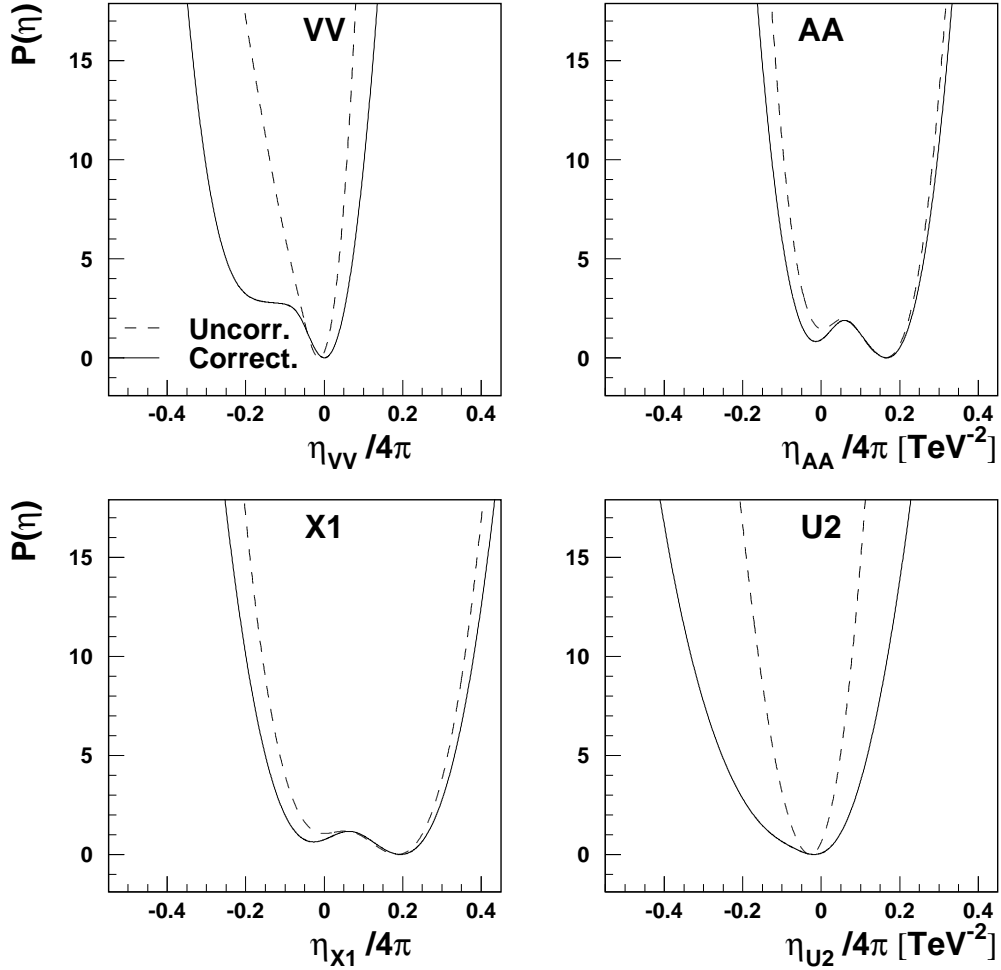}
  }}
  \caption{Log-likelihood function -$\log{\cal P}(\eta)$ for
    ZEUS $e^{+}p$ NC DIS data, for four chosen one-parameter
    models,  as indicated on the plot. The functions are calculated
    with statistical errors only (dashed line) and
    with migration and systematic error corrections (solid line).}
  \label{fig-llsys}
\end{figure}

The influence of the systematic errors and the introduced $Q^{2}$ smearing
on the model probability function ${\cal P}(\vec{\eta})$ has been
studied for the ZEUS $e^{+}p$ NC DIS data \cite{zeus98}.
The results, in terms of  the log-likelihood function, -$\log{\cal P}$,
for four chosen one-parameter models, are shown in Figure \ref{fig-llsys}. 
The applied corrections (mainly the systematic error correction) can have
sizable influence on the model probability distribution. 
Taking into account statistical errors only leads 
to much narrower probability distribution and gives 
much stronger constraints.
The most prominent effect is observed for the VV model.
A narrow probability maximum (minimum of  -$\log{\cal P}$ function)
observed when only statistical errors are included,
becomes wider with a ``shoulder''  on one side when
the systematic errors are taken into account.

\begin{figure}[tp]
\centerline{\resizebox{\figwidth}{!}{%
  \includegraphics{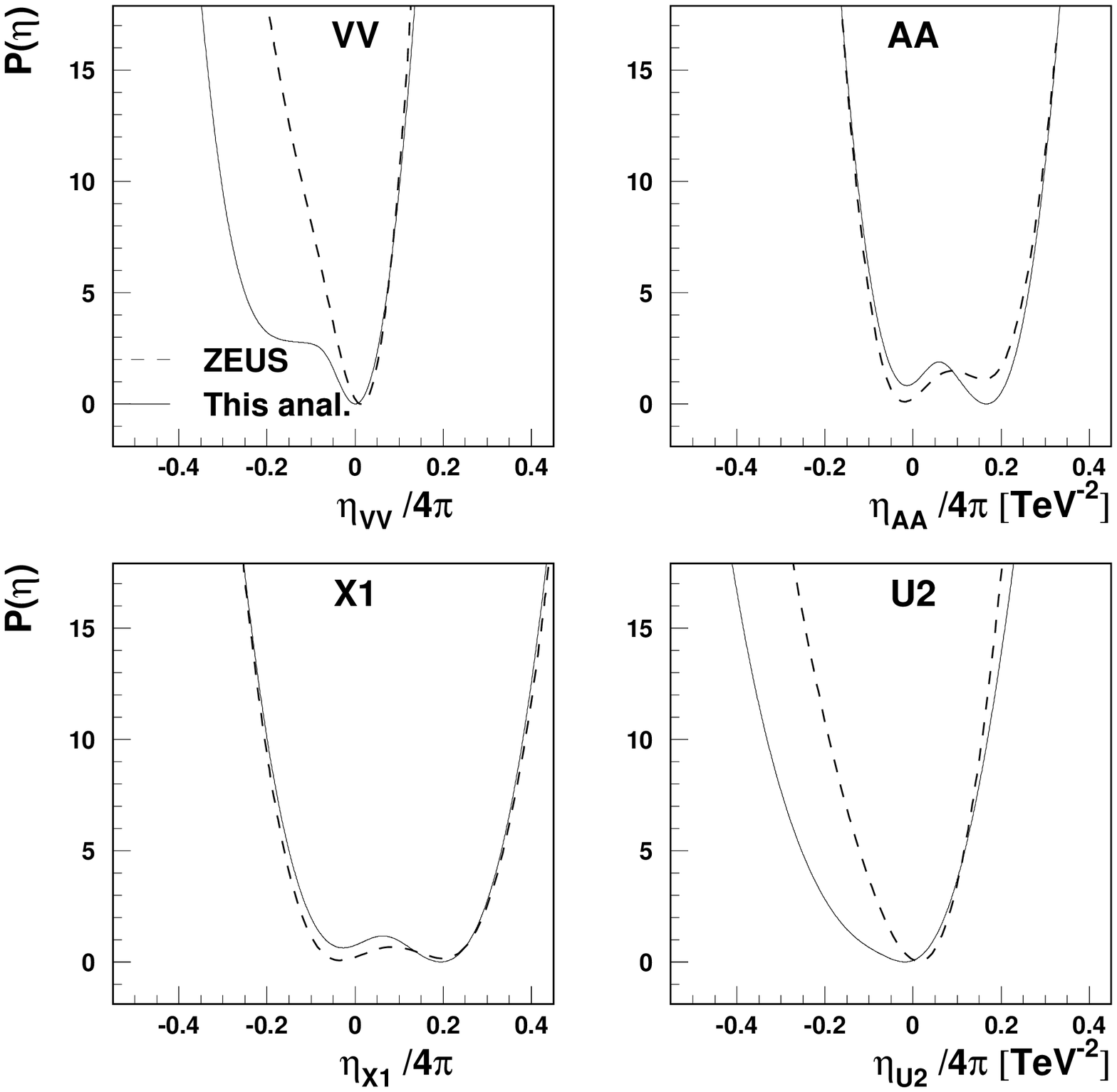}
}}
  \caption{Log-likelihood function -$\log{\cal P}(\eta)$ for
    ZEUS $e^{+}p$ NC DIS data, for four chosen one-parameter
    models, as indicated on the plot.
    The results from this analysis (solid line) are compared with
    the ZEUS results obtained with full detector simulation (dashed line).
    }
  \label{fig-llcomp}
\end{figure}

The results from this analysis have been compared 
with the ZEUS results based on full detector simulation \cite{zeusci}. 
The comparison for the same four one-parameter models is presented 
in Figure \ref{fig-llcomp}.
For some models very good agreement is observed
between this analysis and ZEUS results, as can be seen for AA and X1 models. 
%
% This convinces us that both systematic errors and  event migrations 
% between $Q^{2}$ bins are correctly take into account.
%
However, for models such as VV or U2, the constraints given by ZEUS are
stronger (probability distribution narrower) than the constraints resulting 
from this analysis.  
This is due to the fact that the ZEUS analysis takes into account 
the two-dimensional event distribution in the $(x,y)$ plane, 
whereas this analysis uses the one-dimensional $Q^{2}$ distribution 
only. 

\subsection{Radiative corrections}

For high-energy data from HERA, LEP and the Tevatron, Standard Model
predictions given by experiments are used to rescale leading-order
expectations of the contact interaction models (see formula
\eqref{eq-neta} and \eqref{eq-seta}). 
This accounts not only for different experimental effects, but also for
higher order QCD and electroweak corrections, including radiative
corrections. 
This approach is reasonable as long as the difference
between the corrections for the Standard Model and for the model
including contact interactions is negligible.
It is natural to assume that this difference should be much
smaller than the correction itself.

The contribution of radiative corrections to high-Q$^{2}$ DIS at HERA is
of the order of 10\%.
For high-mass Drell-Yan lepton pair production at the Tevatron 
it is only about 6\%.
Therefore, the possible variation of the radiative corrections for both
HERA and Tevatron data have been neglected.
The only data where radiative corrections could be significant 
is the hadronic cross-section measurement at LEP. 

Most of the events observed at LEP2 are radiative events.
This is due to the "radiative escape" to the $Z^{\circ}$ peak.
Radiation probability 
is significantly enhanced as the
$e^{+}e^{-}$ annihilation cross-section at $\sqrt{s}=M_{Z}$ 
is several orders of magnitude higher than at nominal $\sqrt{s}$.
The leading order cross-section \eqref{eq-lep} is corrected for
radiation effects using the formula \cite{leprad}
\begin{eqnarray}
\sigma_{rad}(s, \vec{\eta} ) & = & 
\bigint{s'_{min}}{s} \frac{ds'}{s} \;\;\;
G(\frac{s'}{s}) \cdot \sigma^{LO} (s' ,  \vec{\eta})  \nonumber
\end{eqnarray}
where integration runs over the center-of-mass energy squared $s'$
of the produced quark pair, and  $s'_{min}$ is the minimum value of
$s'$ required by the event selection cuts\footnote{Data used in this
analysis correspond to $\sqrt{s'/s}>0.9$ (ALEPH) or $\sqrt{s'/s}>0.85$
(DELPHI, L3 and OPAL). This choice significantly reduces possible
influence of radiative corrections.}.  
G(z) is the "radiator function" encapsulating the results of QED
virtual and real corrections.
Used in this analysis is the approximate formula (based on 
 \cite{leprad,leprad2} )
\begin{eqnarray}
G(z) & = & 
f_{r} \cdot \beta (1-z)^{\beta - 1} \; + \; (1-f_{r}) \cdot \delta (1-z)
 \nonumber \\[3mm]
{\rm where } \;\;\;\;\;\;\;\;
 \beta & = & 2 \; \frac{\alpha_{em}}{\pi} \;  
  \left( log \frac{s}{m_{e}^{2}} - 1 \right) \nonumber
\end{eqnarray}
The parameter $f_{r}$ is chosen to reproduce
the cross-section  ratio for radiative and non-radiative 
events\footnote{Events with 0.1$<\sqrt{s'/s}<$0.85 and 
 $\sqrt{s'/s}>$0.85 \cite{opal}.}.

It turned out that the effect of radiative corrections on the probability
function ${\cal P}(\vec{\eta})$ is very small.
The resulting limits on contact interaction mass parameters decrease 
by at most 3\%.

\subsection{Probability functions}
\label{sec-met-prob}

The probability function ${\cal P}(\vec{\eta})$ summarises our current
experimental knowledge about possible $eeqq$ contact interactions.
It will be used to set limits on contact interaction mass scale
parameters and to extract predictions concerning possible future
discoveries.
It is therefore very important to understand the precise meaning of
${\cal P}(\vec{\eta})$.

${\cal P}(\vec{\eta})$ is {\bf not} a probability distribution of
$\vec{\eta}$. A probability distribution should describe the probability of
finding a given value of variable. Our situation is different. Function
${\cal P}(\vec{\eta})$ describes the probability that our data come from
the model described by the set of couplings $\vec{\eta}$ (see section
\ref{sec-met-stat}). It is our data set, which is a variable, and
$\vec{\eta}$ is a set of model parameters: they are unknown, but they
are fixed.
This simple observation has very important implications for this
analysis, not only for the limit setting procedure (see next
subsection) but also for calculation of model predictions.

To set limits on possible deviations from the Standard Model
predictions (eg. for NC DIS cross-section at very high-Q$^{2}$ at HERA
or for hadronic cross-section at next $e^{+}e^{-}$ collider),
we have to consider the probability function $P(r)$, where
the cross-section deviation $r$ is defined as
\begin{eqnarray}
r & = & \frac{\sigma(\vec{\eta})}{\sigma_{SM}} \; = \; R(\vec{\eta})
\nonumber
\end{eqnarray}
If ${\cal P}(\vec{\eta})$ is taken as probability distribution, then
the probability distribution for $r$ should be calculated as
\begin{eqnarray}
P(r)  & = & 
\int d^{N}\vec{\eta} \;\;\; {\cal P}(\vec{\eta}) \; \delta(r-R(\vec{\eta}))  
\label{eq-oldprob}
\end{eqnarray}
where integration is performed over N-dimensional coupling space.
This however leads to completely false results, as is demonstrated 
in appendix \ref{app-toy}.
Instead of calculating the probability distribution for $r$ (which is
not well defined), we should rather try to find out what is
the probability that our data come from the model predicting 
deviation $r$. 
This leads to the formula:
\begin{eqnarray}
P(r)  & = & 
 \bigmod{\left<{\cal P}(\vec{\eta}) \right>}{R(\vec{\eta})=r}  
\nonumber
\end{eqnarray}
where averaging is necessary, if we want to reduce number of parameters
of the probability function (for multi-parameter models).
The commonly used assumption in that case, is that $\vec{\eta}$ has flat
underlying (prior) distribution\footnote{This corresponds to the
assumption, that we would have no preferences for any value of
$\vec{\eta}$, if there is no experimental data.}.
The formula for $P(r)$ can be then expressed as
\begin{eqnarray}
P(r)  & = & \frac{ \displaystyle
\int d^{N}\vec{\eta} \; {\cal P}^{2}(\vec{\eta}) \; \delta(r-R(\vec{\eta})) 
}{ \displaystyle
\int d^{N}\vec{\eta} \; {\cal P}(\vec{\eta}) \; \delta(r-R(\vec{\eta})) 
} \label{eq-newprob}
\end{eqnarray}
The formula applies for any variable which can be used as a parameter of the
probability function. 
In this analysis it will also be used to calculate probability
functions and to set limits on mass scale
parameters corresponding to single couplings in multi-parameter models.

As ${\cal P}(\vec{\eta})$ is not the probability distribution it does
not satisfy any normalisation condition.
Instead it is convenient to rescale the probability function in such a
way that its global maximum has the value of 1:
\begin{eqnarray}
\max_{\vec{\eta}} {\cal P}(\vec{\eta}) & = &  1.  \label{eq-pmax}
\end{eqnarray}

\subsection{Extracting limits}
\label{sec-met-lim}

After imposing condition \eqref{eq-pmax}, the lower and upper limits on the
value of the model parameter $r$ are defined as minimum ($r^{-}$) and
maximum ($r^{+}$ ) values satisfying relation
\begin{eqnarray}
P(r^{-}) & = &  0.05  \nonumber \\
{\rm and} \;\;\;\;\;
P(r^{+}) & = &  0.05 \; . \nonumber
\end{eqnarray}
For any model described by the parameter $r < r^{-}$ or $r > r^{+}$,
the probability that our data results from this model is less than 5\% of
the maximum probability. 
This is taken as the definition of the 95\% confidence level (CL) limits.

For one-parameter contact interaction models this approach is slightly
modified. As models with negative and positive values of $\eta$ are
usually considered as independent scenarios (differing by the signs of the
interference terms in the cross-section), the upper and lower limits on
$\eta$ are calculated using restricted $\eta$ range:
\begin{eqnarray}
P(\eta^{-}) & = &  0.05 \cdot \max_{\eta < 0} P(\eta)  \nonumber \\
{\rm and} \;\;\;\;\;
P(\eta^{+}) & = &  0.05 \cdot \max_{\eta > 0} P(\eta) \; . \label{eq-elimit}
\end{eqnarray}

For one-parameter contact interaction models, or for probability
functions related to single couplings in multi-parameter models, the limits
on coupling values $\eta^{-}$ and $\eta^{+}$  can be translated 
into the limits on contact interaction mass scales
\begin{eqnarray}
\Lambda^{-} & = &  \sqrt{\frac{4 \pi}{ - \eta^{-}}}  \nonumber \\
\Lambda^{+} & = &  \sqrt{\frac{4 \pi}{\eta^{+}}}  \nonumber
\end{eqnarray}

Mass limits commonly used in literature are based on $\eta$ limits
defined in a slightly different way. 
In this paper they will be denoted as $\eta^{--}$
and $\eta^{++}$. Their definition follows from the equations:
\begin{eqnarray}
\bigint{\eta^{--}}{0} d\eta \; P(\eta) & = &  
0.95 \cdot \bigint{-\infty}{0} d\eta \; P(\eta) \nonumber \\
\bigint{0}{\eta^{++}} d\eta \; P(\eta) & = &  
0.95 \cdot \bigint{0}{\infty} d\eta \; P(\eta) \label{eq-elimit2}
\end{eqnarray}
This approach is based on the assumption that $\eta$ has a flat
underlying (prior) distribution.
In such a case $P(\eta)$ can be treated as the probability distribution
for $\eta$.
The mass scale limits corresponding to $\eta^{--}$ and $\eta^{++}$ will
be denoted as $\Lambda^{--}$ and $\Lambda^{++}$.
 Although the definition resulting from equation \eqref{eq-elimit} 
is considered to be more appropriate for this
analysis than definition \eqref{eq-elimit2}, the results for both 
definitions are presented to allow comparison with other results.

As definitions \eqref{eq-elimit} and \eqref{eq-elimit2} correspond to
the different interpretation of the probability function, they are not
expected to give similar results.
In fact, the allowed range for parameter $\eta$, calculated with
equation \eqref{eq-elimit} is usually about 25\% wider than the one
calculated with equation \eqref{eq-elimit2}\footnote{For Gaussian shape
of the probability function, $\eta^{-}$ and $\eta^{+}$ correspond
to $\pm 2.45 \sigma$ limits, whereas $\eta^{--}$ and $\eta^{++}$ to
$\pm 1.96 \sigma$.}.
As a result, corresponding mass scale limits $\Lambda^{-}$ and
$\Lambda^{+}$ are usually 10 to 15\% smaller than $\Lambda^{--}$ and
$\Lambda^{++}$. 

%---------------------------------------------------------------------------

\section{Results}
\label{sec-results}

For one-parameter models the analysis has been performed both without 
and with the additional $SU(2)_{L} \times U(1)_{Y}$
universality assumption (see section \ref{sec-model}).
In the latter case
data coming from neutrino-nucleus scattering experiments and
from different Charged Current processes ( refer section \ref{sec-data})
have been also used to constraint contact interaction couplings.
In the following,
the models assuming $SU(2)_{L} \times U(1)_{Y}$ symmetry are
referred to as SU(2) models.
One-parameter models without  $SU(2)_{L} \times U(1)_{Y}$ 
symmetry will be referred to as simple models, to avoid 
possible confusion.
For all one-parameter models quark and lepton family universality 
is assumed.

Using the overall model probability ${\cal P}(\vec{\eta})$,
as defined by equation \eqref{eq-prob}, the "best" values of
contact interaction couplings (i.e. corresponding to the 
maximum probability) were found using the MINUIT package \cite{minuit}.
The results for one-parameter simple and SU(2) models
are presented in Tables \ref{tab-result1f} and \ref{tab-result1t},
respectively. 
The errors attributed to  $\eta$ values correspond to the decrease in 
the model probability ${\cal P}(\eta)$ by the factor of $\sqrt{e}$.
In case of asymmetric errors the arithmetic mean is given.

\begin{table}[tp]
  \begin{center}
    \begin{tabular}{cr@{~$\pm$}lrrrr}
      \hline\hline\hline\noalign{\smallskip}
Model & \multicolumn{2}{c}{$\eta$} 
      & \multicolumn{4}{c}{Mass scale limits  [TeV]} \\ 
\cline{4-7}\noalign{\smallskip}
      & \multicolumn{2}{c}{[TeV$^{-2}$]} &
 $\Lambda^{-}$ & $\Lambda^{+}$ & $\Lambda^{--}$ & $\Lambda^{++}$ \\ 
\hline\hline\hline\noalign{\smallskip}
 VV &  -0.015 &  0.049 &  9.8 & 10.7 & 11.0 & 11.8 \\
 AA &   0.007 &  0.048 & 10.5 & 10.1 & 11.7 & 11.3 \\
 VA &   0.049 &  0.143 &  6.6 &  6.2 &  7.3 &  6.9 \\
 \hline\noalign{\smallskip}
 X1 &   0.014 &  0.073 &  8.7 &  8.1 &  9.6 &  9.2 \\
 X2 &  -0.011 &  0.075 &  8.2 &  8.4 &  9.2 &  9.4 \\
 X3 &  -0.003 &  0.051 &  9.9 & 10.2 & 11.1 & 11.4 \\
 X4 &  -0.113 &  0.138 &  5.7 &  5.2 &  6.4 &  5.3 \\
 X5 &  -0.079 &  0.132 &  5.9 &  6.4 &  6.6 &  7.0 \\
 X6 &  -0.013 &  0.147 &  6.2 &  5.8 &  7.0 &  6.4 \\
 \hline\noalign{\smallskip}
 U1 &  -0.059 &  0.104 &  6.4 &  7.7 &  7.3 &  8.4 \\
 U2 &  -0.065 &  0.082 &  6.9 &  9.1 &  7.8 &  9.9 \\
 U3 &  -0.044 &  0.053 &  8.5 & 11.7 &  9.6 & 12.7 \\
 U4 &  -0.136 &  0.166 &  5.1 &  5.5 &  5.7 &  5.8 \\
 U5 &  -0.093 &  0.092 &  6.4 &  8.8 &  7.2 &  9.5 \\
 U6 &   0.115 &  0.128 &  7.0 &  5.6 &  7.4 &  6.3 \\
\hline\hline\hline\noalign{\smallskip}
    \end{tabular}
  \end{center}
  \caption{Coupling values and 95\% CL
      mass scale limits resulting from fits of one-parameter models
      {\bf without} $SU(2)_{L} \times U(1)_{Y}$  universality.
The errors attributed to  $\eta$ values correspond to the decrease in 
the model probability ${\cal P}(\eta)$ by the factor of $\sqrt{e}$.
      See text for explanation of the symbols. }
  \label{tab-result1f}
\end{table}

\begin{figure}[tp]
\centerline{\resizebox{\figwidth}{!}{%
  \includegraphics{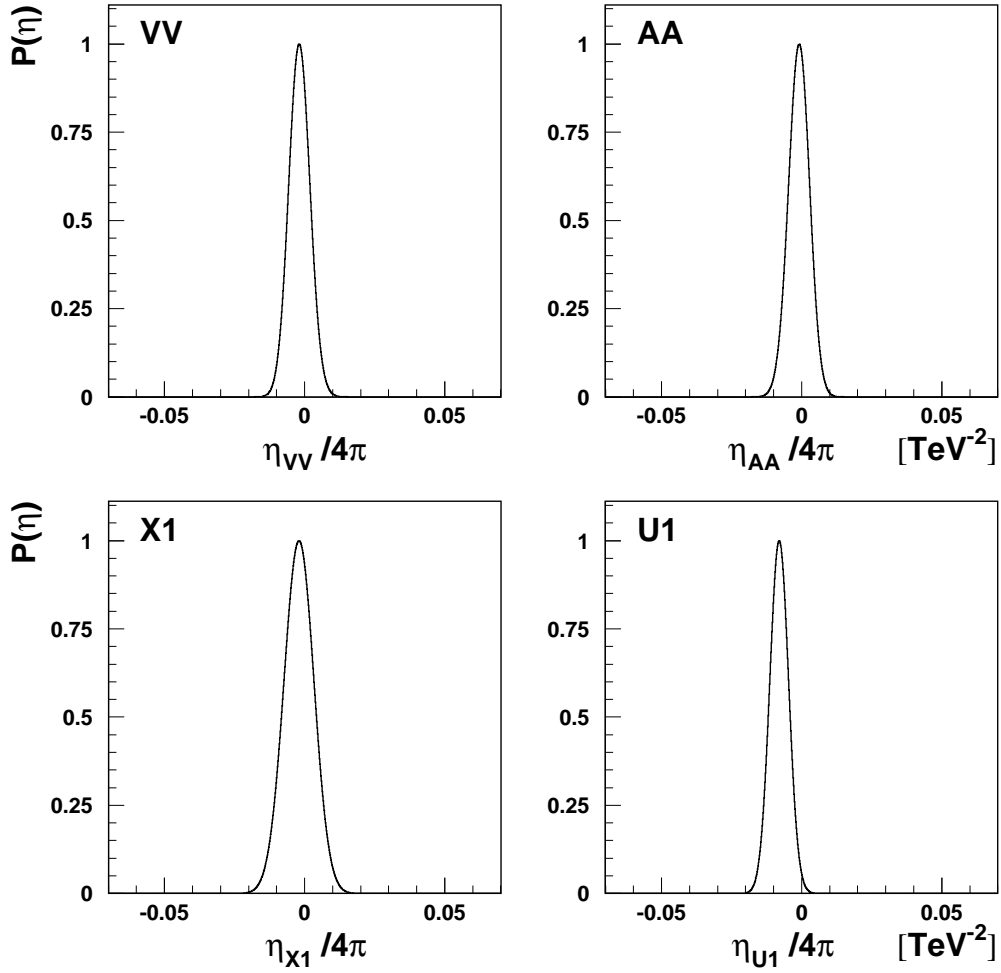}
}}
  \caption{ Probability functions ${\cal P}(\eta)$ for chosen 
    one-parameter models with $SU(2)_{L} \times U(1)_{Y}$  universality, 
    as indicated on the plot.}
  \label{fig-prob1}
\end{figure}

For all simple one-parameter models considered couplings are found to be 
consistent with the Standard Model within $1 \sigma$.
%                                         =====
The same is true for most SU(2) models.
However, for the SU(2) models U1 and U3  the ``best'' coupling values 
are more than $2 \sigma$ from the Standard Model. 
%             =====
These are the only two models which allow for
$\eta^{CC} \neq 0$ (i.e. $\eta^{eu}_{LL} \neq \eta^{ed}_{LL} $). 
The observed deviation from the Standard Model predictions is 
directly  related to  $\eta^{CC}$ bounds coming from the unitarity 
of the CKM matrix and  the $e$-$\mu$ universality, as described in 
section \ref{sec-data}. 
However, it has to be noticed that other data also do support this effect:
the discrepancy observed for the combined data is more significant than
for the Charged Current sector only.
Although the effect is interesting, the data are still in acceptable
agreement with the Standard Model. 
The probability that our data result from the Standard Model equals
  5.7\% and 7.0\% for the U1 and U3 SU(2) models respectively.
% =====     =====
Assuming that there is no direct evidence for $eeqq$ contact interactions,
the limits on single couplings can be calculated. 
The results on mass scale limits  $\Lambda^{-}$, $\Lambda^{+}$,
$\Lambda^{--}$ and $\Lambda^{++}$ obtained from fitting one-parameter 
models are summarised in Tables \ref{tab-result1f} 
and \ref{tab-result1t}. 
For simple models mass limits range
from  5.1 TeV ($\Lambda^{-}$ for the U4 model) 
%     =====            =====        =====
to   11.7 TeV ($\Lambda^{+}$ for the U3 model).
%     =====            =====        =====
Similar limits are obtained for most of the SU(2) models.
Only for the U1 and U3 SU(2) models much higher $\Lambda^{+}$ limits 
of 17.0 and 18.2 TeV are obtained.
%  =====   =====
The probability functions ${\cal P}(\eta)$ for four selected SU(2) models 
are shown in Figure \ref{fig-prob1}.

\begin{table}[tp]
  \begin{center}
    \begin{tabular}{cr@{~$\pm$}lrrrr}
      \hline\hline\hline\noalign{\smallskip}
Model & \multicolumn{2}{c}{$\eta$} 
      & \multicolumn{4}{c}{Mass scale limits  [TeV]} \\
 \cline{4-7}\noalign{\smallskip}
      & \multicolumn{2}{c}{[TeV$^{-2}$]} &
 $\Lambda^{-}$ & $\Lambda^{+}$ & $\Lambda^{--}$ & $\Lambda^{++}$ \\ 
 \hline\hline\hline\noalign{\smallskip}
 VV &  -0.024 &  0.047 &  9.6 & 11.4 & 10.8 & 12.5 \\
 AA &  -0.010 &  0.047 &  9.9 & 11.1 & 11.1 & 12.3 \\
 VA &  -0.078 &  0.108 &  6.3 &  8.0 &  7.1 &  8.7 \\
 \hline\noalign{\smallskip}
 X1 &  -0.025 &  0.067 &  8.1 &  9.5 &  9.2 & 10.5 \\
 X2 &  -0.041 &  0.069 &  7.8 &  9.6 &  8.8 & 10.5 \\
 X3 &  -0.019 &  0.049 &  9.5 & 11.1 & 10.7 & 12.2 \\
 X4 &  -0.066 &  0.144 &  6.0 &  5.4 &  6.7 &  5.8 \\
 X5 &  -0.040 &  0.131 &  6.2 &  6.4 &  7.0 &  7.0 \\
 X6 &  -0.013 &  0.147 &  6.2 &  5.8 &  7.0 &  6.4 \\
 \hline\noalign{\smallskip}
 U1 &  -0.100 &  0.042 &  7.9 & 17.0 &  8.6 & 17.8 \\
 U3 &  -0.083 &  0.036 &  8.6 & 18.2 &  9.4 & 19.1 \\
 U5 &  -0.050 &  0.082 &  7.1 &  8.8 &  8.0 &  9.6 \\
\hline\hline\hline\noalign{\smallskip}
    \end{tabular}
  \end{center}
  \caption{Coupling values and 
      mass scale limits resulting from fits of one-parameter models
      {\bf with} $SU(2)_{L} \times U(1)_{Y}$  universality.
The errors attributed to  $\eta$ values correspond to the decrease in 
the model probability ${\cal P}(\eta)$ by the factor of $\sqrt{e}$.
      See text for explanation of the symbols. }
  \label{tab-result1t}
\end{table}

%
% Contributions of different data to the fit: requested by referee
%

Contribution of different data sets to the mass scale limits presented in
Tables \ref{tab-result1f} and \ref{tab-result1t} can be estimated 
using the probability function.
Mass scale limits $\Lambda^{-}$ and  $\Lambda^{+}$, 
derived from coupling limits $\eta^{-}$ and $\eta^{+}$,
correspond to the decrease of the global probability  ${\cal P}(\eta)$ 
to 0.05 of its maximum value (see equation \eqref{eq-elimit}).
This decrease can be presented as a product of contributions
from all data samples.
Table \ref{tab-probdata} presents the relative probability changes,
calculated separately for different data sets, 
corresponding to mass scale limits $\Lambda^{-}$ and  $\Lambda^{+}$,
for different one-parameter SU(2) models.
The product of numbers in every row is equal to the factor 0.05
defining the 95\% CL.
Numbers close to 1.0 demonstrate that given data set has negligible
influence on the considered mass scale limit.
The smaller the number, the more sensitive are the data to a given CI model.
Numbers greater than 1.0 indicate that the model with mass scale 
 $\Lambda^{-}$ or  $\Lambda^{+}$ gives better description of given data set 
than the ``best'' coupling value resulting from the combined fit.

The results presented in Table  \ref{tab-probdata} show that 
in most models the strongest constraints on
contact interaction couplings come from LEP data on 
forward-backward asymmetries $A^{q}_{FB}$
and on quark production ratios $R_{q}$.
However, for particular models a significant contribution can result
from LEP hadronic cross-section measurements,
neutrino-nucleus scattering data,
HERA NC DIS data or from data on Charged Current interactions.

%
% Table of probability contributions
%

\renewcommand{\multirowsetup}{\centering}
\begin{table}[tp]
   \begin{tabular*}{\textwidth}{@{}cc@{\extracolsep{4in minus 4in}}cccccccc@{}}
      \hline\hline\hline\noalign{\smallskip}
 & Mass & \multicolumn{8}{c}{Relative change in model probability} \\
\cline{3-10}\noalign{\smallskip}
Model & scale  & HERA & Tevatron & 
\multicolumn{3}{c}{LEP} & 
\multicolumn{2}{c}{Low energy NC} &  
CC \\ \cline{5-7} \cline{8-9}\noalign{\smallskip}
  & limit & $e^{\pm}p$ NC  & Drell-Yan  &
$\sigma_{had}$ & $R_{q}$ &  $A^{q}_{FB}$ &
$l^{\pm}N$ & $\nu N$ & data \\
\hline\hline\hline\noalign{\smallskip}
\multirow{2}{10mm}{VV} & $\Lambda^{-}$ & 
0.697 & 0.887 & 
1.344 & 1.665 & 0.028 & 
1.000 & 1.300 &   \\
   & $\Lambda^{+}$ & 
1.107 & 0.632 & 
0.751 & 0.205 & 0.841 & 
1.000 & 0.553 &   \\
\noalign{\smallskip}
\multirow{2}{10mm}{AA} & $\Lambda^{-}$ & 
1.240 & 1.021 & 
0.536 & 1.867 & 0.020 & 
0.984 & 1.961 &   \\
   & $\Lambda^{+}$ & 
0.774 & 0.835 & 
1.451 & 0.208 & 0.722 & 
1.013 & 0.350 &   \\
\noalign{\smallskip}
\multirow{2}{10mm}{VA} & $\Lambda^{-}$ & 
1.264 & 0.710 & 
1.281 & 1.092 & 0.033 & 
0.916 & 1.316 &   \\
   & $\Lambda^{+}$ & 
1.008 & 0.928 & 
1.370 & 0.811 & 0.504 & 
1.059 & 0.090 &   \\
\hline\noalign{\smallskip}
\multirow{2}{10mm}{X1} & $\Lambda^{-}$ & 
1.214 & 1.013 & 
0.694 & 1.746 & 0.017 & 
0.962 & 2.060 &   \\
   & $\Lambda^{+}$ & 
0.830 & 0.885 & 
1.395 & 0.344 & 0.634 & 
1.031 & 0.217 &   \\
\noalign{\smallskip}
\multirow{2}{10mm}{X2} & $\Lambda^{-}$ & 
0.842 & 0.931 & 
1.323 & 1.659 & 0.017 & 
0.973 & 1.773 &   \\
   & $\Lambda^{+}$ & 
1.079 & 0.748 & 
0.878 & 0.327 & 0.660 & 
1.021 & 0.320 &   \\
\noalign{\smallskip}
\multirow{2}{10mm}{X3} & $\Lambda^{-}$ & 
1.016 & 1.005 & 
0.993 & 1.734 & 0.017 & 
0.992 & 1.720 &   \\
   & $\Lambda^{+}$ & 
0.977 & 0.710 & 
1.137 & 0.192 & 0.738 & 
1.007 & 0.444 &   \\
\noalign{\smallskip}
\multirow{2}{10mm}{X4} & $\Lambda^{-}$ & 
0.331 & 0.762 & 
0.318 & 1.021 & 1.414 & 
1.018 & 0.424 &   \\
   & $\Lambda^{+}$ & 
0.814 & 0.666 & 
0.218 & 0.779 & 0.690 & 
0.969 & 0.813 &   \\
\noalign{\smallskip}
\multirow{2}{10mm}{X5} & $\Lambda^{-}$ & 
0.617 & 0.650 & 
1.178 & 1.749 & 0.133 & 
1.043 & 0.435 &   \\
   & $\Lambda^{+}$ & 
1.142 & 0.516 & 
0.884 & 0.064 & 1.411 & 
0.949 & 1.114 &   \\
\noalign{\smallskip}
\multirow{2}{10mm}{X6} & $\Lambda^{-}$ & 
0.771 & 0.786 & 
1.127 & 0.077 & 0.984 & 
0.965 &  &  \\
   & $\Lambda^{+}$ & 
1.367 & 0.731 & 
0.789 & 1.976 & 0.031 & 
1.025 &  &  \\
\hline\noalign{\smallskip}
\multirow{2}{10mm}{U1} & $\Lambda^{-}$ & 
1.141 & 0.963 & 
0.895 & 1.207 & 0.361 & 
0.925 & 0.443 & 0.285  \\
   & $\Lambda^{+}$ & 
0.933 & 0.954 & 
0.753 & 0.903 & 1.174 & 
1.028 & 0.323 & 0.212  \\
\noalign{\smallskip}
\multirow{2}{10mm}{U3} & $\Lambda^{-}$ & 
1.056 & 0.855 & 
0.254 & 1.289 & 0.243 & 
0.987 & 1.157 & 0.611  \\
   & $\Lambda^{+}$ & 
0.963 & 0.885 & 
0.502 & 0.804 & 1.212 & 
1.005 & 0.449 & 0.266  \\
\noalign{\smallskip}
\multirow{2}{10mm}{U5} & $\Lambda^{-}$ & 
0.724 & 0.849 & 
0.447 & 1.535 & 0.561 & 
1.084 & 0.195 &   \\
   & $\Lambda^{+}$ & 
1.004 & 0.671 & 
0.109 & 0.570 & 1.267 & 
0.910 & 1.036 &   \\
\hline\hline\hline\noalign{\smallskip}
    \end{tabular*}
  \caption{Relative changes in model probabilities calculated
           for separate data sets (as indicated in the table) 
           corresponding to the decrease in the global probability 
           at mass scale limit ($\Lambda^{-}$ or  $\Lambda^{+}$) 
           to 0.05 of its maximum value (for negative or positive
           couplings respectively). 
           Considered are one-parameter models with family and
            $SU(2)_{L} \times U(1)_{Y}$ universality.
           }
  \label{tab-probdata}
\end{table}

\begin{figure}[tp]
\centerline{\resizebox{\figwidth}{!}{%
  \includegraphics{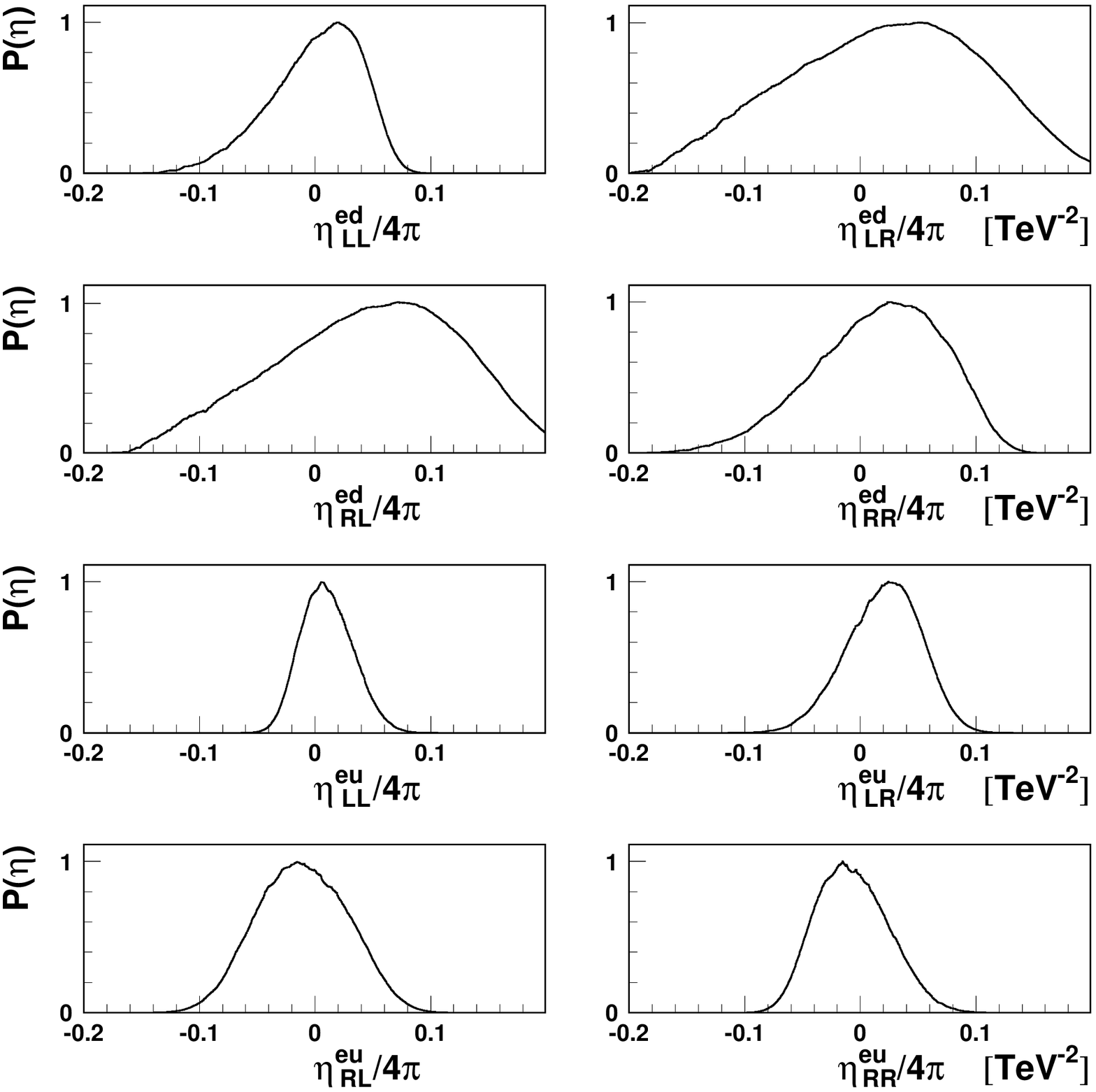}
}}
  \caption{ Probability functions $P(\eta)$ for single
    contact interaction couplings (as indicated on the plot) obtained
    within the general contact interaction model.}
  \label{fig-prob2}
\end{figure}

%
% Moved figure to the begining !
%
The probability functions
for single couplings obtained for the general model are presented 
in Figure \ref{fig-prob2}. All couplings are consistent with the 
Standard Model predictions.
Results for single couplings obtained for the general model, the model
with family universality and the SU(2) model with family universality
are summarised in Table \ref{tab-result2}. It has to be stressed that
all limits for single couplings are derived without any assumptions 
concerning the remaining couplings, 
which corresponds to the definition \eqref{eq-newprob} of
a probability function. For this reason most calculated limits are
weaker than in case of one-parameter models. The mass limits
obtained for the general model range 
from 2.1 TeV ($\Lambda^{ed\;+}_{RL}$)
%   =====       =====      
to 5.1 TeV ($\Lambda^{eu\;-}_{LL}$). 
%  =====              ======
For the SU(2) model with family universality,
the corresponding numbers are 3.5 and 7.8 TeV 
%                            =====   =====
for $\Lambda^{ed\;-}_{RR}$ and $\Lambda^{eu\;+}_{LL}$, respectively. 
%             =====                      =====

\begin{table*}[tp]
   \begin{tabular*}{\textwidth}{c@{\extracolsep{4in minus 4in}}rlrlrlrlrlrl}
      \hline\hline\hline\noalign{\smallskip}
         & \multicolumn{12}{c}{Mass scale limits [TeV] }  \\ 
\cline{2-13}\noalign{\smallskip}
         & \multicolumn{4}{c}{ } &
           \multicolumn{4}{c}{Model with} &
           \multicolumn{4}{c}{SU(2) model with}  \\
Coupling & \multicolumn{4}{c}{General model } &
           \multicolumn{4}{c}{family universality} &
           \multicolumn{4}{c}{family universality}  \\ 
                  \cline{2-5} \cline{6-9} \cline{10-13} \noalign{\smallskip}
 &  $\Lambda^{-}$ & $\Lambda^{+}$ & $\Lambda^{--}$ & $\Lambda^{++}$
 &  $\Lambda^{-}$ & $\Lambda^{+}$ & $\Lambda^{--}$ & $\Lambda^{++}$
 &  $\Lambda^{-}$ & $\Lambda^{+}$ & $\Lambda^{--}$ & $\Lambda^{++}$ \\ 
\hline\hline\hline\noalign{\smallskip}
 $\eta^{ed}_{LL}$ &
 3.1 & 3.6 &  3.4 &  4.0 &
 4.3 & 5.4 &  4.8 &  6.0 &
 7.5 & 6.1 &  8.1 &  6.8 \\
 $\eta^{ed}_{LR}$ &
 2.4 & 2.2 &  2.6 &  2.5 &
 2.8 & 3.2 &  3.0 &  3.6 &
 3.6 & 3.7 &  4.0 &  4.1 \\
 $\eta^{ed}_{RL}$ &
 2.6 & 2.1 &  2.8 &  2.4 &
 3.9 & 2.8 &  4.4 &  3.1 &
 4.5 & 3.9 &  5.0 &  4.4 \\
 $\eta^{ed}_{RR}$ &
 2.8 & 2.8 &  3.0 &  3.1 &
 3.3 & 3.5 &  3.7 &  3.9 &
 3.5 & 5.1 &  3.8 &  5.7 \\
 $\eta^{eu}_{LL}$ &
 5.1 & 3.9 &  5.6 &  4.3 &
 5.3 & 4.1 &  5.9 &  4.5 &
 6.5 & 7.8 &  7.3 &  8.5 \\
 $\eta^{eu}_{LR}$ &
 4.1 & 3.3 &  4.3 &  3.7 &
 3.7 & 3.5 &  4.1 &  3.9 &
 4.9 & 4.7 &  5.4 &  5.3 \\
 $\eta^{eu}_{RL}$ &
 3.1 & 3.5 &  3.5 &  3.9 &
 3.1 & 3.5 &  3.4 &  3.9 &
 4.5 & 3.9 &  5.0 &  4.4 \\
 $\eta^{eu}_{RR}$ &
 3.7 & 3.7 &  4.1 &  4.1 &
 3.8 & 4.2 &  4.3 &  4.6 &
 4.8 & 4.4 &  5.4 &  4.8 \\
\hline\hline\hline\noalign{\smallskip}
    \end{tabular*}
  \caption{95\% CL mass scale limits for single couplings, obtained within
        different models, as indicated in the table. See text for 
        mass scale limits definition.}
  \label{tab-result2}
\end{table*}

\begin{figure}[tp]
\centerline{\resizebox{\figwidth}{!}{%
  \includegraphics{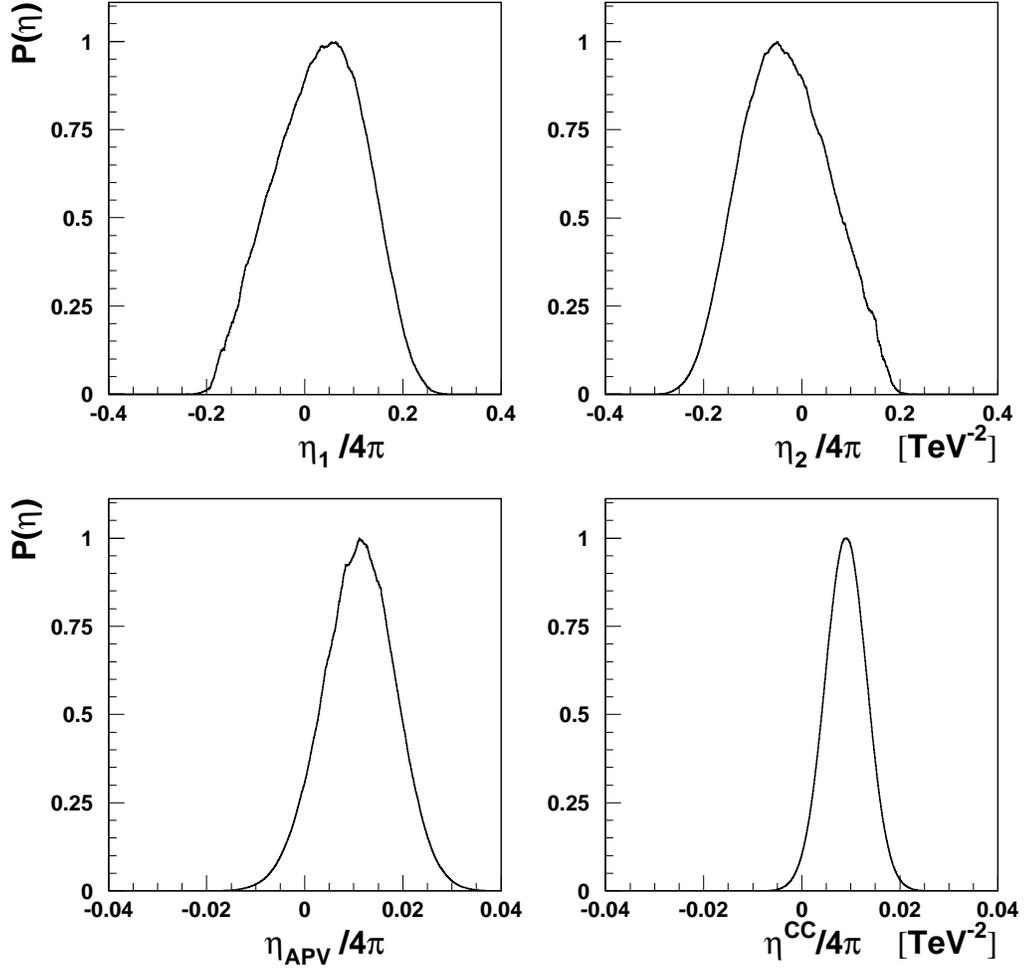}
}}
  \caption{ Probability functions, calculated within the general 
       contact interaction model, for the two least constrained coupling 
       combinations $\eta_{1}$ and  $\eta_{2}$ (upper plots),
       the atomic parity violating combination $\eta_{APV}$ (lower left) 
       and the CC contact interaction coupling $\eta^{CC}$ induced 
       in the SU(2) model (lower right plot).
       Note different horizontal scales between upper and lower plots.}
  \label{fig-prob3}
\end{figure}

Single couplings can either increase or decrease cross-section for a
given process, as compared with the Standard Model expectations.
It is therefore also possible that the influence of the two different
couplings compensate each other. 
Because of that, the limits on mass scales
obtained for single couplings does not exclude contact interactions
with smaller mass scales.  To obtain the most general limit,
the eigenvectors of the correlation matrix (obtained from MINUIT 
from the functional form of ${\cal P}(\vec{\eta})$ in the vicinity
of the maximum probability) are considered. 
In case of the general model the two least constrained linear coupling
combinations are
\begin{eqnarray}
\eta_{1} & = &  -0.26 \eta^{ed}_{LL} + 0.84 \eta^{ed}_{LR} 
                +0.15 \eta^{ed}_{RL} + 0.33 \eta^{ed}_{RR}  \nonumber \\
           &  & -0.06 \eta^{eu}_{LL} + 0.11 \eta^{eu}_{LR} 
                -0.10 \eta^{eu}_{RL} + 0.27 \eta^{eu}_{RR} \nonumber \\
%               ==========================================
{\rm and} \;\;\;\;\;\;\;
\eta_{2} & = &  +0.20 \eta^{ed}_{LL} + 0.17 \eta^{ed}_{LR} 
                +0.81 \eta^{ed}_{RL} - 0.48 \eta^{ed}_{RR}  \nonumber \\
           &  & +0.10 \eta^{eu}_{LL} - 0.14 \eta^{eu}_{LR} 
                +0.10 \eta^{eu}_{RL} - 0.09 \eta^{eu}_{RR} \nonumber
%               ==========================================
\end{eqnarray}
This is in agreement with the observation that the least constrained
single couplings are $\eta^{ed}_{LR}$ and $\eta^{ed}_{RL}$, which can
also be seen  from Table \ref{tab-result2}.
The probability functions for $\eta_{1}$ and $\eta_{2}$ are shown in 
Figure \ref{fig-prob3}. The mass scale limit corresponding to $\eta_{1}$
is\footnote{As the sign of $\eta_{1}$ is arbitrary, only one value is
given. It is calculated as min($\Lambda^{+}_{1},\Lambda^{-}_{1}$).} 
\begin{eqnarray}
\Lambda_{1} & = & 2.1 \; \rm TeV  \nonumber 
%                =====
\end{eqnarray}
This limit should be considered to be the most general one, as it
is valid for any combination of couplings. 
This means that any contact interaction with a mass scale below 2.1 TeV
%                                                               =====
is excluded on 95\% CL. On the other hand it also shows that 
the existing data do not exclude mass scales of the order
of 3 TeV.
% ======
%
The limits on the mass scale associated with $\eta_{1}$  are summarised
in Table \ref{tab-result3}.

\begin{table*}[tp]
   \begin{tabular*}{\textwidth}{c@{\extracolsep{4in minus 4in}}rlrlrlrlrlrl}
      \hline\hline\hline\noalign{\smallskip}
         & \multicolumn{12}{c}{Mass scale limits [TeV] }  \\ 
\cline{2-13}\noalign{\smallskip}
         & \multicolumn{4}{c}{ } &
           \multicolumn{4}{c}{Model with} &
           \multicolumn{4}{c}{SU(2) model with}  \\
Coupling & \multicolumn{4}{c}{General model } &
           \multicolumn{4}{c}{family universality} &
           \multicolumn{4}{c}{family universality}  \\ 
   \cline{2-5} \cline{6-9} \cline{10-13} \noalign{\smallskip}
 &  $\Lambda^{-}$ & $\Lambda^{+}$ & $\Lambda^{--}$ & $\Lambda^{++}$
 &  $\Lambda^{-}$ & $\Lambda^{+}$ & $\Lambda^{--}$ & $\Lambda^{++}$
 &  $\Lambda^{-}$ & $\Lambda^{+}$ & $\Lambda^{--}$ & $\Lambda^{++}$ \\ 
\hline\hline\hline\noalign{\smallskip}
 $\eta_{1}$ &
 \multicolumn{2}{c}{2.1}  & \multicolumn{2}{c}{2.3} &
 \multicolumn{2}{c}{2.7}  & \multicolumn{2}{c}{3.0} &
 \multicolumn{2}{c}{3.1}  & \multicolumn{2}{c}{3.5} \\
 $\eta_{APV}$ &
 9.8 & 6.0 &  10.4 &  6.6 &
 9.7 & 6.1 &  10.3 &  6.7 &
 10.9 & 7.5 &  11.7 &  8.4 \\
 $\eta^{ed}_{LL} - \eta^{eu}_{LL}$ &
 2.8 & 3.5 &  3.1 &  3.9 &
 3.4 & 4.3 &  3.8 &  4.8 &
 14.4 & 7.2 &  15.1 &  7.9 \\
\hline\hline\hline\noalign{\smallskip}
    \end{tabular*}
  \caption{95\% CL  mass-scale limits 
           corresponding to the least constrained 
           coupling combination $\eta_{1}$, atomic parity 
           violating coupling combination $\eta_{APV}$ 
           and  $\eta^{ed}_{LL}-\eta^{eu}_{LL}$ combination
           corresponding to Charged Current contact interaction
           coupling $\eta^{CC}$ of SU(2) model.
           As the sign of $\eta_{1}$ is arbitrary, only one value is
           given.}
  \label{tab-result3}
\end{table*}

Shown in the same table are mass limits corresponding to the atomic 
parity violating combination of couplings, 

\begin{eqnarray}
\eta_{APV} & \equiv &  \eta^{ed}_{LL} + \eta^{ed}_{LR} 
                 - \eta^{ed}_{RL} - \eta^{ed}_{RR}  \nonumber \\
           & + &  \eta^{eu}_{LL} + \eta^{eu}_{LR} -
                \eta^{eu}_{RL} - \eta^{eu}_{RR} \nonumber
\end{eqnarray}
$\eta_{APV}$ is close to the most strongly constrained coupling combination
(the eigenvector with the highest eigenvalue). Mass scale limits up to
about 11~TeV are obtained. The probability function for $\eta_{APV}$ 
%     =====
in the case of the general model is included in Figure \ref{fig-prob3}. 

Also shown in Figure \ref{fig-prob3} is the probability function 
for the Charged Current contact interaction coupling $\eta^{CC}$ 
induced in the SU(2) model.
The discrepancy between the data and the Standard Model has decreased slightly,
as compared with the U1 and U3 SU(2) models.
The most probable value of $\eta^{CC}$  is about $2\sigma$ from 
the Standard Model value, which corresponds to 
the probability of about 10\%.
%                   ======
%
This discrepancy is observed for the SU(2) model only.
When $SU(2)_{L} \times U(1)_{Y}$  universality is not assumed 
(i.e. in case of the general model and the model with family 
universality) the corresponding coupling combination is no longer related
to the Charged Current sector and is in good agreement with 
the Standard Model.
The corresponding mass scale limits are included in Table \ref{tab-result3}.

%---------------------------------------------------------------------------

\section{Predictions}
\label{sec-predictions}

All presented results are in good agreement with the Standard
Model. Nevertheless, an interesting question is whether 
"new physics" in terms of contact interactions can be expected to show up 
in high-energy experiments in the near future. 

\begin{figure}[tp]
\centerline{\resizebox{!}{\figheight}{%
  \includegraphics{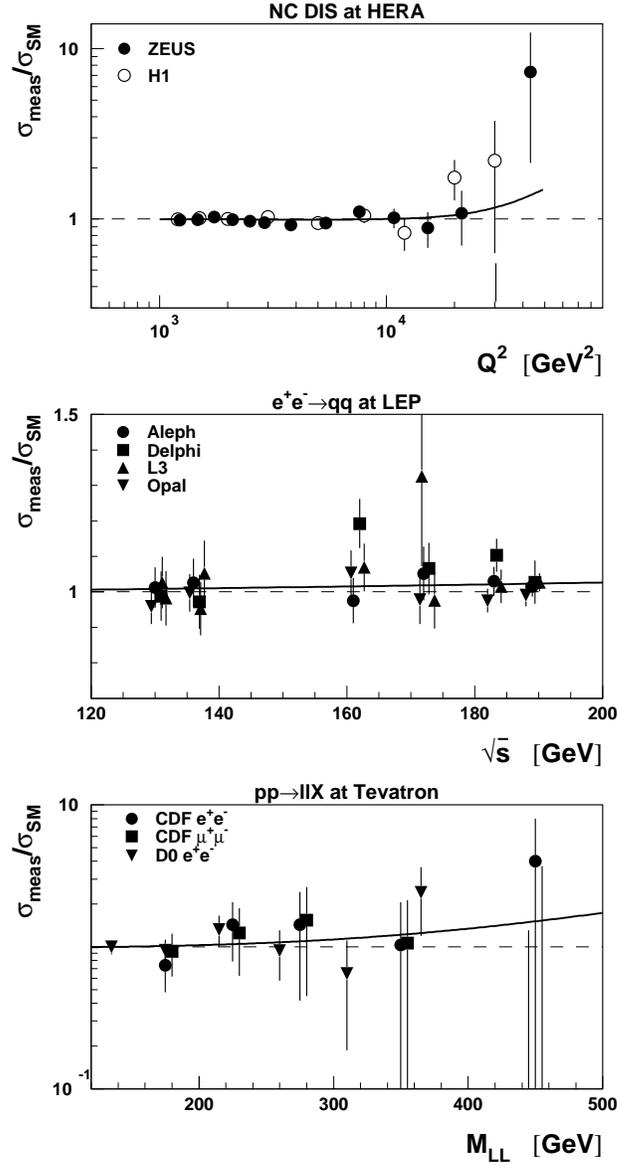}
}}
  \caption{Cross-section deviations from the Standard Model
           resulting from the general contact interaction model fit
           (thick solid line) compared with HERA, LEP and the Tevatron data.}
  \label{fig-compare}
\end{figure}

The cross-sections corresponding to the "best fit" of the general model 
(the set of coupling values resulting in the best description of all data, 
i.e. corresponding to the maximum probability)
are compared in Figure \ref{fig-compare} with the HERA, 
LEP and Tevatron data. 
In the case of LEP data, the best fit of the general model agrees very well
with the Standard Model. The Contact Interaction contribution to
the measured cross-section does not exceed 3\% for $\sqrt{s}$
up to 200 GeV. 
On the other hand, the same model predicts for both HERA and the Tevatron 
an increase in the cross-section by almost a factor of 2
at the highest $Q^{2}$/$M_{ll}$.
In order to verify the significance of these predictions it is
unavoidable to  consider the statistical uncertainty of these predictions.

\begin{figure}[tp]
\centerline{\resizebox{\figwidth}{!}{%
  \includegraphics{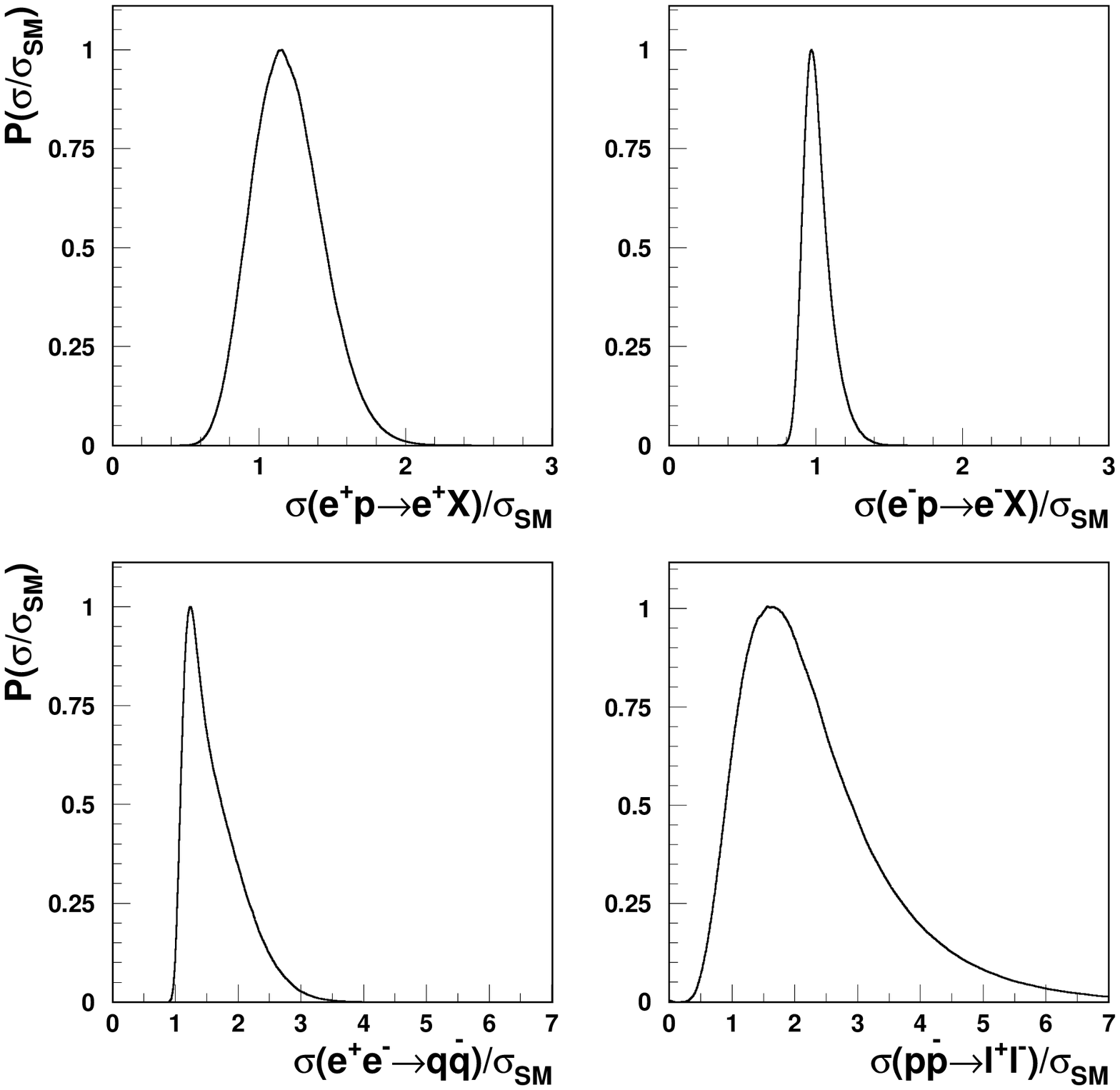}
}}
  \caption{Probability functions for possible deviations from 
  the Standard Model predictions for: $e^{+}p$ and $e^{-}p$ NC DIS
cross-section at HERA, at $Q^{2}$ = 30,000 GeV$^{2}$ (upper plots), 
$e^{+}e^{-}$ total hadronic cross-section at $\sqrt{s}$ = 400 GeV
(lower left plot) and Drell-Yan lepton pair production cross-section at
the Tevatron, at $M_{ll}$ = 500 GeV (lower right plot).}
  \label{fig-predhist}
\end{figure}

Employing the Monte Carlo techniques,
the probability function for the contact interaction 
couplings, ${\cal P}(\vec{\eta})$, is translated into 
the probability function for relevant cross-section deviations, as
described in section \ref{sec-met-prob}.
Considered in this analysis are possible deviations from 
the Standard Model predictions for high-$Q^{2}$ $e^{-}p$ and $e^{+}p$ 
scattering at HERA\footnote{For the proton beam energy of 920 GeV and 
the electron/positron beam energy of 27.5 GeV.} (see section \ref{sec-hera}),
for the total quark pair production cross-section at LEP 
(or Next Linear Collider, NLC; see section \ref{sec-lep}) 
and for the Drell-Yan lepton pair production
at the Tevatron  (see section \ref{sec-dy}).
The probability functions calculated for these processes at selected 
energy scales are presented in Figure \ref{fig-predhist}.

\begin{figure}[tp]
\centerline{\resizebox{\figwidth}{!}{%
  \includegraphics{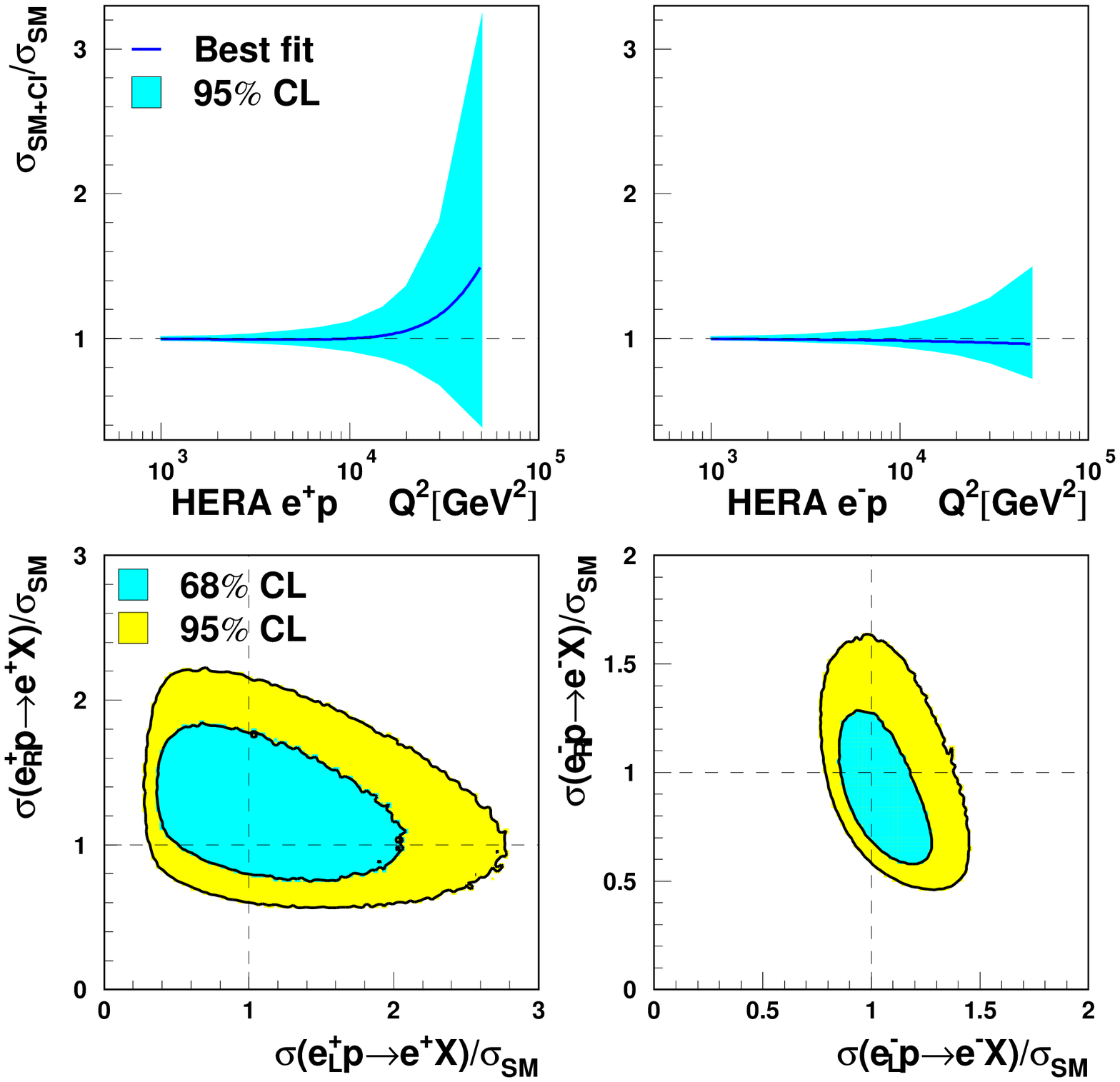}
}}
  \caption{The 95\% CL limit band on the ratio of predicted
to the Standard Model cross-section for  $e^{+}p$ and $e^{-}p$ NC DIS
scattering at HERA (upper plots) and the 68\% and 95\% CL contours for
the possible deviations for scattering of right- and left-handed
electrons and positrons at $Q^{2}=$30,000 GeV$^{2}$ (lower plots).
The limits are calculated using the general contact interaction  model.}
  \label{fig-predhera1}
\end{figure}

\begin{figure}[tp]
\centerline{\resizebox{\figwidth}{!}{%
  \includegraphics{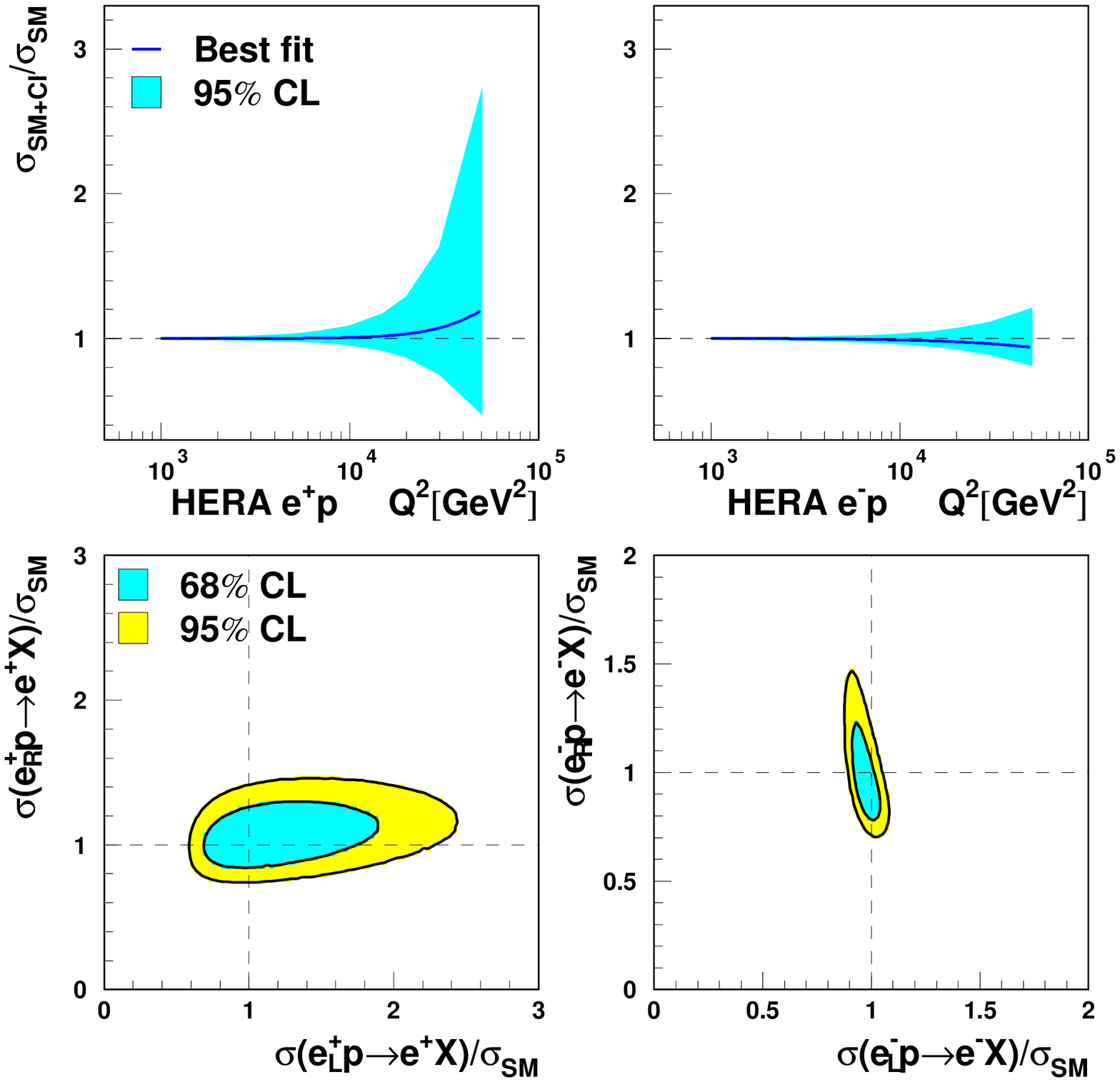}
}}
  \caption{The 95\% CL limit band on the ratio of predicted
to the Standard Model cross-section for  $e^{+}p$ and $e^{-}p$ NC DIS
scattering at HERA (upper plots) and the 68\% and 95\% CL contours for
the possible deviations for scattering of right- and left-handed
electrons and positrons at $Q^{2}=$30,000 GeV$^{2}$ (lower plots).
The limits are calculated using the SU(2) contact interaction  model with
family universality.}
  \label{fig-predhera2}
\end{figure}

The results for HERA, in terms of the 95\% confidence limit bands on the
ratio of predicted and the Standard Model cross-sections as a function
of Q$^{2}$, are shown  in Figures \ref{fig-predhera1} and \ref{fig-predhera2}
for the general model and the SU(2) model with family universality,
respectively. 
For the $e^{+}p$ NC DIS the uncertainty of these
predictions is very big, although the nominal predictions
of both models are above the Standard Model. 
The Standard Model prediction is well within the 95\% confidence level band. 
For the general model, the increase in the $e^{+}p$ NC DIS cross-section at 
HERA by up to about 80\% at $Q^{2}$ of 30,000 GeV$^{2}$ would still be 
%                  =====
consistent with current experimental data.
For the SU(2) model the corresponding limit is 63\%.
%                                             =====
It turns out that the best statistical sensitivity (in single measurement)
to possible contact interaction effects is obtained when considering
the number of events measured for $Q^{2} \; >$ 15,000 GeV$^{2}$.
The allowed increase in the integrated $e^{+}p$ NC DIS cross-section 
is about 40\% for the general model and about 30\% for the SU(2) model.
%       =====                                =====
%
In order to reach the level of statistical precision,
which would allow them to confirm possible discrepancy of this 
size\footnote{We require that the allowed increase in the 
cross-section for $Q^{2} >$ 15,000 GeV$^{2}$ (at 95\% CL) should 
correspond to at least three times the statistical error on the number
of events. 5\% systematic uncertainty on the expected number of events
is assumed.}, HERA experiments would have to collect $e^{+}p$
luminosities of the order of 100-200 pb$^{-1}$ (depending on the model). 
%                             =====
This will be possible after the HERA upgrade planned for year 2000.

Constraints on the possible deviations from the Standard Model 
predictions are much stronger in case of $e^{-}p$ NC DIS.
This is because the Standard Model cross-section itself
is higher, and also because different contact interaction coupling
combinations contribute. 
It is interesting to notice that the possible cross-section increase
for $e^{+}p$ NC DIS, which is suggested by global fit results,
corresponds to {\bf decrease} in the NC DIS cross-section for $e^{-}p$.
For the general model, deviations
larger than about 20\% are excluded for $Q^{2} >$15,000 GeV$^{2}$,
%                =====
whereas for the SU(2) model with family universality the limit goes down
to about 7\%. 
%       =====
%
When compared with the predicted statistical precision of the future HERA
data, this indicates that it will be very hard
to detect contact interactions in the future HERA $e^{-}p$ running.
For the general model the required luminosity is 
of the order of  400 pb$^{-1}$.
%               =====

However, the HERA "discovery window" can be visibly enlarged if we consider
scattering of polarised electrons and/or positrons.
The 68\% and 95\% CL contours for the allowed deviations for scattering
of right- and left-handed electrons or positrons are included in
Figures \ref{fig-predhera1} (for general model) and \ref{fig-predhera2}
(for SU(2) model), at $Q^{2}=$30,000 GeV$^{2}$. 
In both cases, the cross-section deviations for $e^{+}_{L}p$ and $e^{-}_{R}p$
scattering are less constrained than in case of $e^{+}_{R}p$ and  $e^{-}_{L}p$,
respectively.
For the general model possible deviations for both left- and right-handed
projectiles are significantly higher than in the unpolarised case.
However, for the  SU(2) model, constraints significantly weaker 
than in the unpolarised case are obtained only for  $e^{+}_{L}p$ 
and $e^{-}_{R}p$ scattering.
In both models deviations of up to about 50\%
%                                        =====
are still allowed for $e^{+}_{L}p$ scattering at  $Q^{2}>$15,000 GeV$^{2}$,
assuming 60\% polarisation of the positron beam.
To observe effects of this size it would be enough to collect
luminosity of the order of 70-80 pb$^{-1}$.
%                          =====
%
For $e^{-}_{R}p$ scattering maximum allowed deviations are 28\% and 19\%, 
%                                                          =====   =====
for the general and SU(2) models respectively. 
It means that with 60\% longitudinal $e^{-}_{R}$ polarisation it would be
possible to observe significant deviations from Standard Model predictions
already for luminosities of the order of 120 pb$^{-1}$ (for the 
%                                       =====
general model).  
Unfortunately, polarisation can result in significantly higher systematic
uncertainties of the Standard Model predictions, which was not
considered here.

Since the only visible inconsistency between data and the Standard Model is
observed in the Charged Current sector (for models assuming 
$SU(2)_{L} \times U(1)_{Y}$  universality), 
the  interesting question is whether
any effect can be observed in high $Q^{2}$ CC DIS at HERA.
It turns out that the possible effect is far beyond the HERA sensitivity.
The ``best'' $\eta^{CC}$ value (resulting from the SU(2) model fit) corresponds
to a {\bf decrease} in the CC DIS cross-section at HERA not greater than 2\% 
within the accessible $Q^{2}$ range, and a decrease exceeding 5\% is excluded
at 95\% CL. 
At the same confidence level, any increase in the cross-section
by a similar amount is excluded.

\begin{figure}[tp]
\centerline{\resizebox{\figwidth}{!}{%
  \includegraphics{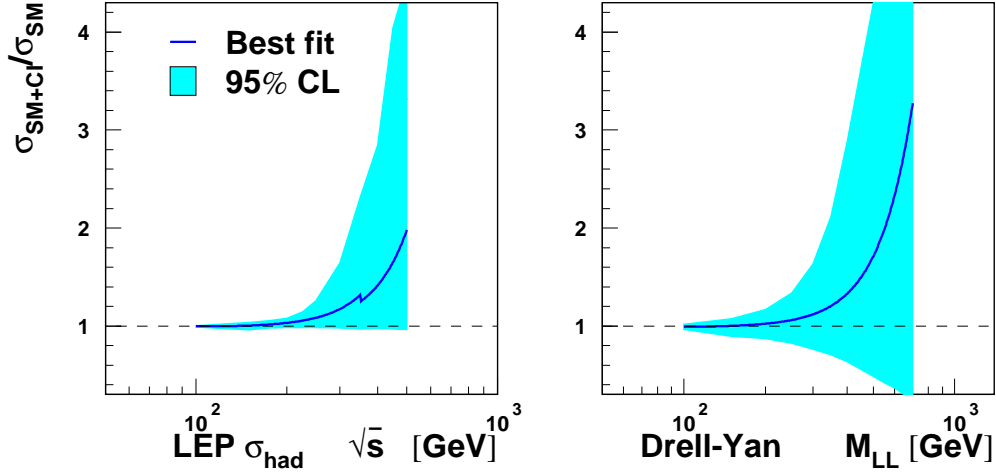}
}}
  \caption{
Left: 
the 95\% CL limit band on the ratio 
$\sigma_{SM+CI}/\sigma_{SM}$, where $\sigma_{SM}$ is the Standard Model
total hadronic cross-section at LEP/NLC and $\sigma_{SM+CI}$ is the
cross-section calculated in the general contact interaction model;
Right: 
the same ratio for the Drell-Yan lepton pair production at the Tevatron.
}
  \label{fig-predlep1}
\end{figure}

\begin{figure}[tp]
\centerline{\resizebox{\figwidth}{!}{%
  \includegraphics{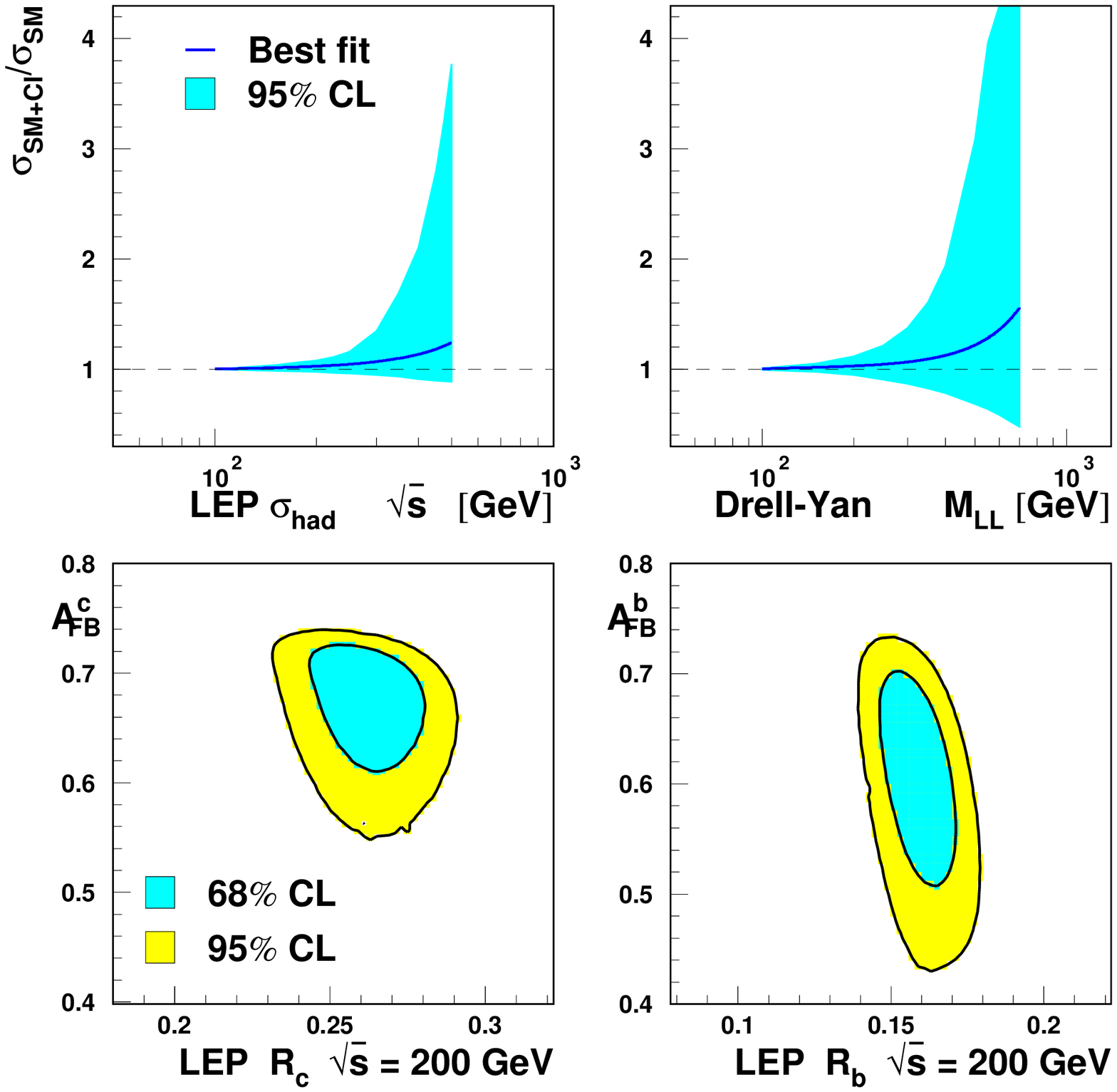}
}}
  \caption{
Upper left: 
the 95\% CL limit band on the ratio 
$\sigma_{SM+CI}/\sigma_{SM}$, where $\sigma_{SM}$ is the Standard Model
total hadronic cross-section at LEP/NLC and $\sigma_{SM+CI}$ is the
cross-section calculated in the SU(2) contact interaction model;
Upper right: 
the same ratio for the Drell-Yan lepton pair production at the Tevatron;
Below:
the 68\% and 95\% CL contours for
the allowed values of the forward-backward asymmetries and the quark
production fractions for $c$ and $b$ quark production at LEP, at
$\sqrt{s}$=200 GeV.
}
  \label{fig-predlep2}
\end{figure}

Model predictions for both the total hadronic cross-section at
electron-positron collider (LEP or NLC) and Drell-Yan lepton pair
production cross-section at the Tevatron are shown in Figures
\ref{fig-predlep1} and \ref{fig-predlep2}, for the general contact
interaction model and the SU(2) model, respectively.
For $e^{-}e^{+} \rightarrow q \bar{q}$ at $\sqrt{s}$ above 
about 300 GeV upper cross-section limits obtained from
both contact interaction models increase rapidly.
Cross-section deviations up to a factor of 3 are allowed for
%                                        =====
$\sqrt{s}\sim$500 GeV. 
Unfortunately, this energy range will be accessible only in the Next
Linear Collider experiment(s).
LEP will not go beyond $\sqrt{s}\sim$200 GeV:
at this energy the possible deviations from the Standard Model are
only about 8\%, which makes possible discovery very difficult.
%         =====

\begin{table}[tp]
  \begin{center}
    \begin{tabular}{lcr@{~-}lr@{~-}lr@{~-}l}
      \hline\hline\hline\noalign{\smallskip}
   & SM  & \multicolumn{6}{c@{}}{Allowed range on 95\% C.L.} \\ 
\cline{3-8}\noalign{\smallskip}
  & Value & \multicolumn{2}{c}{General}   &
             \multicolumn{2}{c}{Model with} & 
             \multicolumn{2}{c@{}}{SU(2) model}  \\
   & (LO) & \multicolumn{2}{c}{model} & 
            \multicolumn{2}{c}{family univ.} & 
            \multicolumn{2}{c@{}}{w. fam. univ.} \\
\hline\hline\hline\noalign{\smallskip}
 $R_{b}$ &  0.159 &
 0.147 & 0.161 &
 0.137 & 0.180 &
 0.139 & 0.179 \\
 $R_{c}$ &  0.262 &
 0.242 & 0.266 &
 0.230 & 0.294 &
 0.232 & 0.291 \\
 $A^{b}_{FB}$ &  0.601 &
 &      &
 0.345 & 0.750 &
 0.431 & 0.732 \\
 $A^{c}_{FB}$ &  0.668 &
 &      &
 0.469 & 0.750 &
 0.551 & 0.738 \\
\hline\hline\hline\noalign{\smallskip}
    \end{tabular}
  \end{center}
  \caption{Leading order Standard Model prediction and the allowed
range (on 95\% CL) for the heavy quark production ratios and
forward-backward asymmetries, for $e^{+}e^{-}$ annihilation at
$\sqrt{s}$=200 GeV, in different contact interaction models.}
  \label{tab-predict2}
\end{table}

However, significant deviations from Standard Model predictions are
still possible for heavy quark production ratios $R_{c}$ and $R_{b}$, and
for the forward-backward asymmetries $A^{c}_{FB}$ and $A^{b}_{FB}$.
The 68\% and 95\% CL contours for the values of
the forward-backward asymmetry versus the quark production fraction,
allowed within the SU(2) model for $c$ and $b$ quark production at
$\sqrt{s}$=200 GeV are included in Figure \ref{fig-predlep2}.
Allowed ranges for  $R_{c}$, $R_{b}$, $A^{c}_{FB}$ and $A^{b}_{FB}$ at
$\sqrt{s}$=200 GeV, for different contact interaction models
considered, are summarised in Table \ref{tab-predict2}.
For the general model, in which contact interactions are limited to
the first quark generation only, variations of $R_{b}$ and $R_{c}$ are
still possible, due to the possible changes in $u\bar{u}$ and
$d\bar{d}$ production cross-sections. However, parton level
forward-backward asymmetries  $A^{c}_{FB}$ and $A^{b}_{FB}$ do not depend
on contact interaction couplings in this model. Therefore limits for
$A^{c}_{FB}$ and $A^{b}_{FB}$  are not reported for the general model.
Heavy quark observables considered here are least constrained for the model
with family universality. Large effects are still possible in this model
for both production fractions and asymmetries. 
Deviations up to about 13\% are possible for $R_{b}$ and $R_{c}$.
%                     ======
Least constrained by the existing experimental data is the
forward-backward asymmetry for the $b\bar{b}$  production $A^{b}_{FB}$,
where deviations from the Standard Model prediction by up to 40\% are
%                                                          ======
still possible. 
Note that for the general model  $R_{b}$ and $R_{c}$ are
100\% correlated, whereas for models with the family universality they
are 100\% anti-correlated.

It seems that the best place to study contact interactions in the
nearest future is the Tevatron, which should run again after being
upgraded in the year 2000. If there is any "new physics" corresponding to
the contact interaction model it is very likely to show up
in Drell-Yan lepton pair production for masses above 200-300 GeV.
Moreover, upper limits on possible deviations from the Standard Model 
predictions are much higher than in case of HERA and LEP/NLC. 
For $M_{ll}=$500 GeV, which should be easily accessible with
increased luminosity, cross-section deviations up to a factor of 5 
%                                                              =====
are still not excluded. 

Upper limits on the cross-section deviations from the Standard
Model predictions, derived on 95\% confidence level
in different contact interaction models are
summarised in Table~\ref{tab-predict}.

\begin{table}[tp]
   \begin{center}
    \begin{tabular}{llrrr}
      \hline\hline\hline\noalign{\smallskip}
   &  & \multicolumn{3}{c}{Limits on $\Delta \sigma / \sigma_{SM}$ [\%]} \\ 
\cline{3-5}\noalign{\smallskip}
Reaction  & Energy & General   &   Model with & SU(2) model \\
          & scale & model & family univ. & w. family univ. \\
\hline\hline\hline\noalign{\smallskip}
 $e^{+}p$ NC DIS     & $Q^{2}$=10000 GeV$^{2}$ &
 11 & 10 & 9 \\
 $\sqrt{s}=$318 GeV  & $Q^{2}$=20000 GeV$^{2}$ &
 36 & 30 & 28 \\
 & $Q^{2}$=30000 GeV$^{2}$ &
 81 & 65 & 63 \\
 & $Q^{2}$=50000 GeV$^{2}$ &
 220 & 180 & 170 \\ \hline\noalign{\smallskip}
 $e^{-}p$ NC DIS     & $Q^{2}$=10000 GeV$^{2}$ &
 8 & 4 & 3 \\
 $\sqrt{s}=$318 GeV  & $Q^{2}$=20000 GeV$^{2}$ &
 18 & 8 & 7 \\
 & $Q^{2}$=30000 GeV$^{2}$ &
 28 & 13 & 11 \\
 & $Q^{2}$=50000 GeV$^{2}$ &
 49 & 26 & 21 \\ \hline\noalign{\smallskip}
 $e^{-}e^{+} \rightarrow q \bar{q}$     & $\sqrt{s}$=175 GeV &
 5 & 5 & 6 \\
 & $\sqrt{s}$=200 GeV &
 8 & 8 & 8 \\
 & $\sqrt{s}$=225 GeV &
 14 & 13 & 11 \\
 & $\sqrt{s}$=250 GeV &
 26 & 24 & 16 \\
 & $\sqrt{s}$=300 GeV &
 65 & 61 & 35 \\
 & $\sqrt{s}$=400 GeV &
 185 & 185 & 110 \\ \hline\noalign{\smallskip}
 $p \bar{p} \rightarrow l^{+} l^{-} X $     & $M_{ll}$=200 GeV &
 17 & 12 & 12 \\
 $\sqrt{s} = $ 1800 GeV & $M_{ll}$=300 GeV &
 64 & 55 & 38 \\
 & $M_{ll}$=400 GeV &
 190 & 185 & 95 \\
 & $M_{ll}$=500 GeV &
 440 & 450 & 210 \\
\hline\hline\hline\noalign{\smallskip}
    \end{tabular}
  \end{center}
  \caption{Upper limits (on 95\% CL) on cross-section deviations from
   the Standard Model predictions in different contact interaction models.
          Considered are $e^{+}p$ and $e^{-}p$ scattering at HERA, 
         total hadronic cross-section at LEP/NLC and Drell-Yan lepton
         pair production at the Tevatron, as indicated in the table.}
  \label{tab-predict}
\end{table}

\begin{figure}[tp]
\centerline{\resizebox{!}{\figheight}{%
  \includegraphics{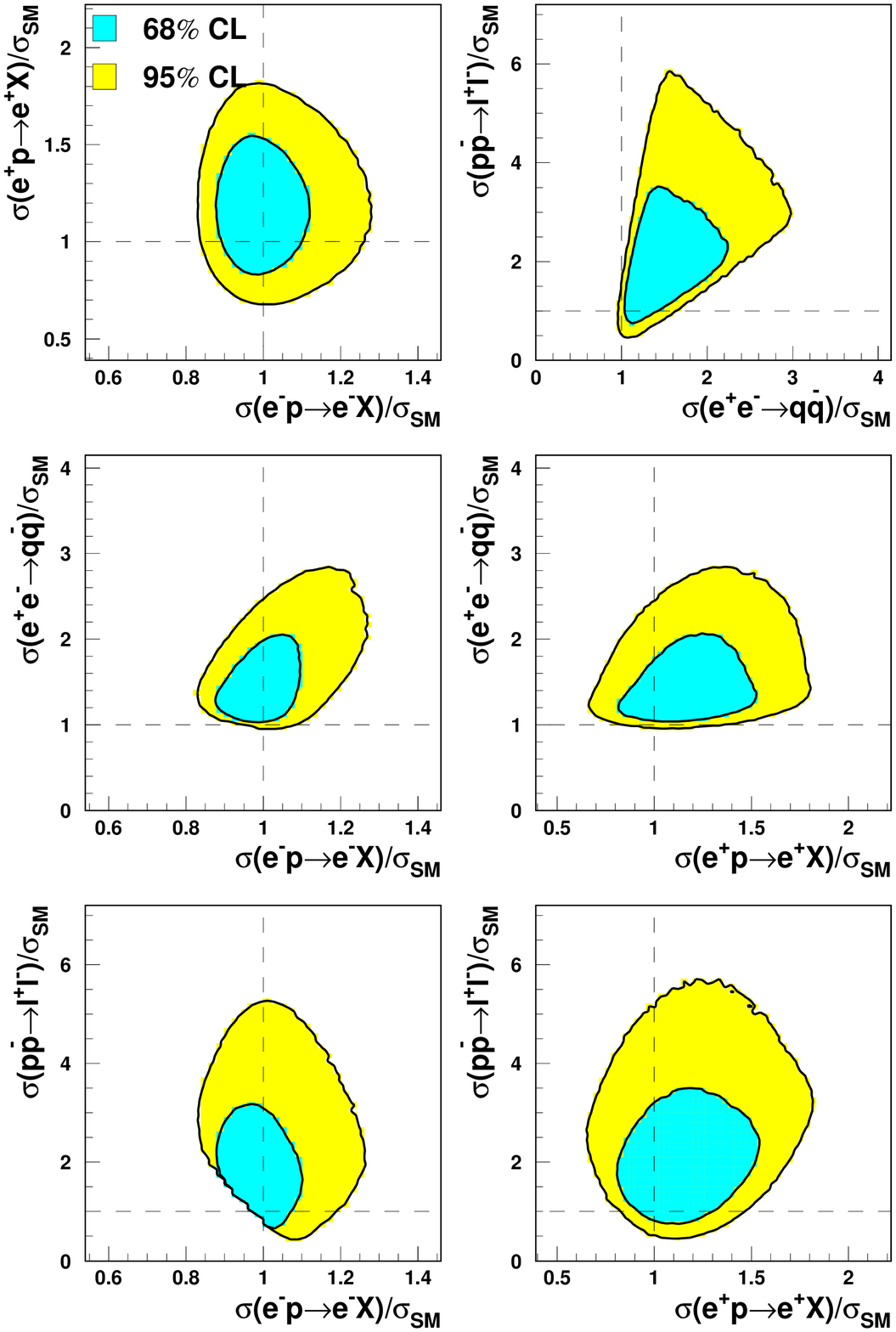}
}}
  \caption{The 68\% and 95\% CL contours for the possible deviation from
the Standard Model predictions, for different combinations of
measurements. Considered are: $e^{+}p$ and $e^{-}p$ NC DIS
cross-section at HERA, at $Q^{2}$ = 30,000 GeV$^{2}$, total
$e^{+}e^{-} \rightarrow q\bar{q}$ cross-section at $\sqrt{s}$ = 400 GeV
and Drell-Yan lepton pair production cross-section at
the Tevatron, at $M_{ll}$ = 500 GeV, as indicated on the plot. 
The limits are calculated using the general contact interaction model.}
  \label{fig-pred2d1}
\end{figure}
\begin{figure}[tp]
\centerline{\resizebox{!}{\figheight}{%
  \includegraphics{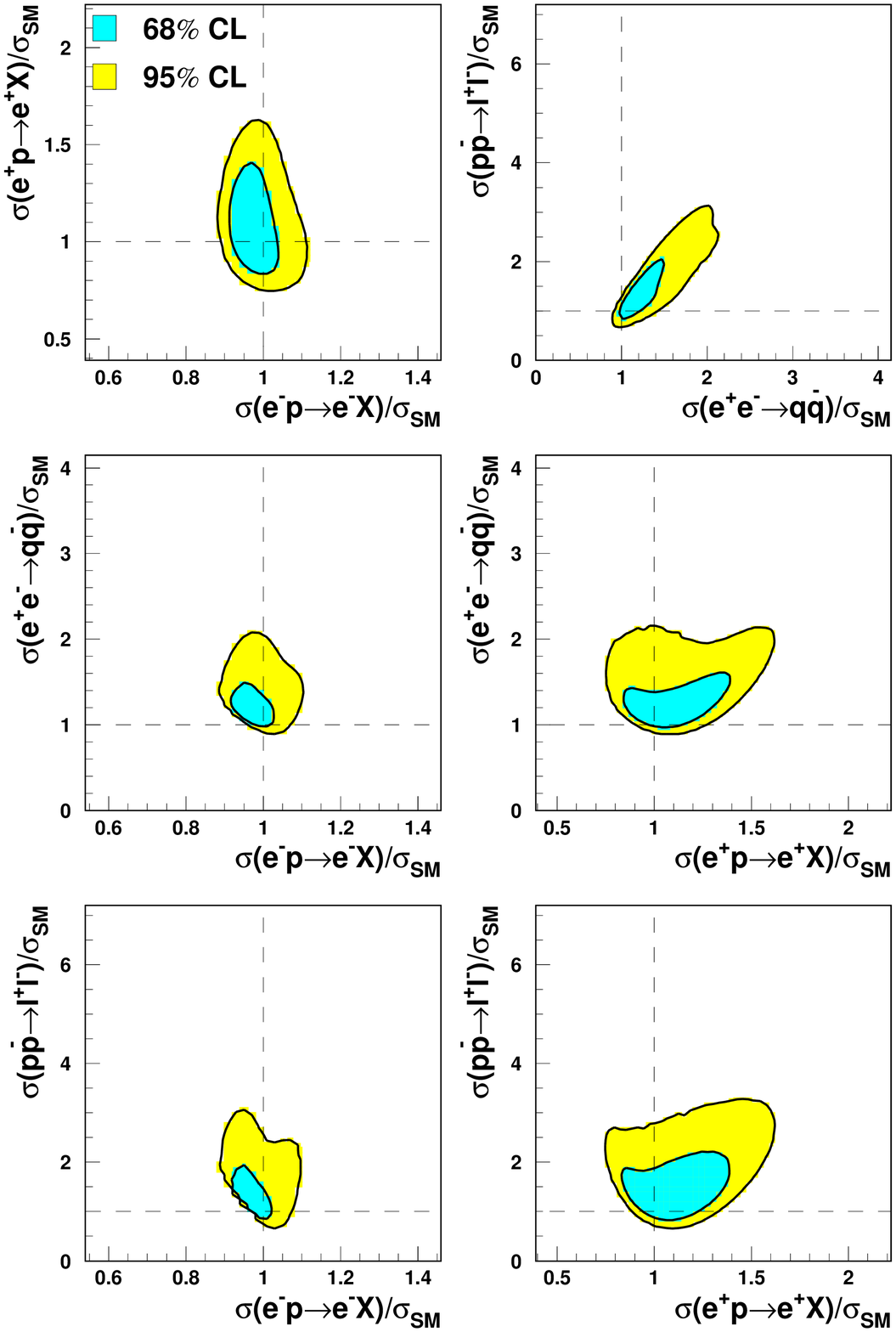}
}}
  \caption{The 68\% and 95\% CL contours for the possible deviation from
the Standard Model predictions, for different combinations of
measurements. Considered are: $e^{+}p$ and $e^{-}p$ NC DIS
cross-section at HERA, at $Q^{2}$ = 30,000 GeV$^{2}$, total
$e^{+}e^{-} \rightarrow q\bar{q}$ cross-section at $\sqrt{s}$ = 400 GeV
and Drell-Yan lepton pair production cross-section at
the Tevatron, at $M_{ll}$ = 500 GeV, as indicated on the plot. 
The limits are calculated using the SU(2) contact interaction
model with family universality.}
  \label{fig-pred2d2}
\end{figure}

When considering possible future discoveries at high-energy
experiments, it is also interesting to study the relation
between effects observed at different experiments.
The 68\% and 95\% CL contours for the sizes of the allowed deviation from
the Standard Model predictions, for different measurement combinations, are
shown in Figures \ref{fig-pred2d1} and \ref{fig-pred2d2}, for the
general contact interaction model and for the SU(2) model with family
universality, respectively.
In both cases, clear correlation is observed between 
the Drell-Yan cross-section deviation at the Tevatron and 
the hadronic $e^{+}e^{-}$ cross-section at LEP/NLC.
Possible cross-section increase at the Tevatron has to be accompanied by 
the increase in the hadronic cross-section at LEP/NLC.
Similar correlation is observed between the hadronic $e^{+}e^{-}$ 
cross-section at LEP/NLC and  $e^{+}p$ NC DIS cross-section at HERA
for SU(2) model.
Another interesting observation is that the possible decrease in
the $e^{-}p$ NC DIS cross-section at HERA should be related to the increase in
both Tevatron and LEP/NLC cross-section.
In other  cases correlations between different measurements are weak.
This shows that contact interaction searches at LEP, the Tevatron and HERA
are, to large extent, independent.
Data from all types of experiments are necessary to constraint contact
interaction model in general case.

%---------------------------------------------------------------------------
\clearpage
\section{Summary}
\label{sec-summary}

Data from HERA, LEP, the Tevatron and low energy experiments 
were used to constrain electron-quark contact interactions. 
The contact interaction mass scale limits obtained for different one-parameter 
models range from 5.1 to about 18 TeV.  
%                =====       =====
Using the most general approach, in which all couplings to
are allowed to vary independently, any contact interactions
with mass scale below 2.1 TeV are  excluded at 95\% CL.
%                    =====
This limit can be raised to 3.1~TeV 
%                          =====
by assuming $SU(2)_{L} \times U(1)_{Y}$ and quark/lepton family 
universality.
There is a slight hint on possible ``new physics''
in the Charged Current sector (related to Neutral Current contact
interactions by $SU(2)_{L} \times U(1)_{Y}$  universality), where
the discrepancy between data and the Standard Model is at the 
$2\sigma$ level.
The mass scale of new Charged Current interactions suggested by the data
is of the order of 10~TeV.
However, this effect - if real - would have negligible impact on predictions
for future collider results.

The limits on possible effects to be observed in future HERA, LEP and 
Tevatron running are estimated. 
Possible deviations from the Standard-Model predictions
for total hadronic cross-section at LEP and $e^{-}p$ scattering 
cross-section at HERA, are already strongly
limited by existing data. 
However, improved experimental sensitivity to new interactions
should result from the measurement of heavy quark production ratios 
and asymmetries at LEP, as well as from polarised electron scattering
at HERA.
Sizable effects are still not excluded for $e^{+}p$ NC DIS at HERA and
the required statistical precision of the data should be accessible
after HERA upgrade.
The best "discovery potential" seems to come from future Tevatron running,
where significant deviations from the Standard Model predictions are 
still allowed. For Drell-Yan lepton pair production cross-section 
deviations at  $M_{ll}$=500 GeV up to a factor of 5 are still not 
%                                               =====
excluded. 
However, all experiments should continue to analyse their data in terms of
possible new electron-quark interactions, as constraints resulting from
different experiments are, to large extent, complementary.

%---------------------------------------------------------------------------
%
\vfill
\section*{Acknowledgements}

I would like to thank all members of the Warsaw HEP group and 
of the ZEUS Collaboration for support, encouragement, many useful
comments and suggestions.
Special thanks are due to U.Katz for very productive discussions 
and many valuable comments to this paper,
and to Prof. A.K.Wr\'{o}blewski for his help in preparing 
the final version of it.
I am also grateful to 
    M.Lancaster and W.Sakumoto from CDF,
    A.Gupta and  A.Kotwal from D0,
    I.Tomalin from {\sc Aleph},
    A.Olchevski from {\sc Delphi},
    F.Filthaut from L3  and
    P.Ward from {\sc Opal}
for their assistance in gathering the relevant experimental data and
in understanding details of the measurements.

This work has been partially supported by the Polish State Committee 
for Scientific Research (grant No. 2 P03B 035 17).

\clearpage

%
%---------------------------------------------------------------------------
%
% Appendix
%

\appendix
\section{Interpretation of the probability function}
\label{app-toy}

In this appendix a simple "toy model" is used to demonstrate
that the probability function, as introduced in section
\ref{sec-method}, should not be treated as the probability distribution
for $\vec{\eta}$.

Let us consider a model with N independent  couplings.
Assume that all data considered in the analysis are in perfect
agreement with the Standard Model and that the resulting probability
function is 
\begin{eqnarray}
{\cal P}(\vec{\eta}) & = & \frac{1}{(\sqrt{2\pi} \sigma)^{N}} \cdot
\exp \left( -\frac{\vec{\eta}^{\;2}}{2 \sigma^{2}} \right) \nonumber
\end{eqnarray}
where $\vec{\eta}^{\;2}  =  \eta_{1}^{2} + ... + \eta_{N}^{2}$ and
the distribution width $\sigma$ is taken to be the same for all couplings.
The Standard Model gives the best description of the data,
corresponding to the maximum value of ${\cal P}(\vec{\eta})$.

Consider the cross-section deviation from the Standard Model
prediction, which is of the form
\begin{eqnarray}
     r(\vec{\eta}) & = & \frac{\sigma(\vec{\eta})}{\sigma_{SM}}
                   \; = \; ( 1 + \vec{\eta}^{\;2})  \nonumber
\end{eqnarray}
If ${\cal P}(\vec{\eta})$ is taken as a probability distribution, then
the probability distribution for $r$ should be calculated from
equation \eqref{eq-oldprob}. 
After integrating over the coupling space we obtain:
\begin{eqnarray}
P(r) & = & 
\frac{(r-1)^{\frac{N}{2}-1}}
     {2^{\frac{N}{2}} \sigma^{N} \Gamma(\frac{N}{2})}
     \cdot \exp \left( -\frac{r-1}{2 \sigma^{2}} \right)  \nonumber
\end{eqnarray}
The shape of $P(r)$ corresponds to that of the
$\chi^{2}$ distribution for N degrees of freedom.
For $N \le 2$, $P(r)$ has a maximum for $r=1$, i.e. for the Standard Model
expectation. 
However, for models with $N \ge 3$ parameters, the maximum of $P(r)$ is
shifted towards $r > 1 $ and the probability of the Standard Model
solution $P(r=1) = 0 $.
This results is incompatible with our initial assumption that
all data are in perfect agreement with the Standard Model.

The above calculation, based on the formula \eqref{eq-oldprob},  is not
correct because it assumes that  ${\cal P}(\vec{\eta})$ is the 
probability distribution for $\vec{\eta}$.
We can treat  ${\cal P}(\vec{\eta})$ as the probability distribution
only if we assume that $\vec{\eta}$ has a flat prior distribution.
This assumption justifies limit setting procedure
described in section \ref{sec-met-lim} (formula \eqref{eq-elimit2}).  
However, it does not justify variable transformation
from $\vec{\eta}$ to $r$ (resulting from equation \eqref{eq-oldprob}), as the
prior distribution for the new variable does not need to be flat.
Instead, one should try to define $P(r)$ in the same way as  
${\cal P}(\vec{\eta})$, i.e. as the probability that our data
come from the model predicting deviation $r$.
This approach results in formula \eqref{eq-newprob}. For our toy
model the probability of observing deviations from the Standard Model
predictions is
\begin{eqnarray}
P(r) & = & \exp \left(-\frac{(r-1)}{2 \sigma^{2}} \right) \nonumber
\end{eqnarray}
where the normalisation condition \eqref{eq-pmax} has been imposed.
The result does not depend on the number of free model parameters
and the most probable model is the one predicting no deviation from the
Standard Model (taking into account that $r \ge 1$).

%---------------------------------------------------------------------------
%

%---------------------------------------------------------------------------

\end{document}